\documentclass[aps,onecolumn,manuscript,amsmath,amssymb,prb]{revtex}
\usepackage{graphicx}% Include figure files
\usepackage{graphics}% Include figure files
\usepackage{color}% Include figure files
\usepackage{rotating}% Include figure files

\def \a{\`{a}}

\def \i{\`{\i}~}

\begin{document}

\title{
Statistics of particle dispersion in
Direct Numerical Simulations of wall-bounded
turbulence: results of an international
collaborative benchmark test
%Collaborative Testing of Lagrangian
%Particle Tracking in Direct Numerical
%Simulations of Wall-Bounded Turbulence
}

\author{
C. Marchioli$^{1,}$\footnote{Corresponding author. E-mail: marchioli@uniud.it, Tel.: +39 0432 558006},
A. Soldati$^{1}$
%C. Marchioli$^1$, A. Soldati$^{1,*}$
%\footnote[]{$^{*}$Author to whom correspondence should be addressed.
%Also affiliated with Department of Fluid
%Mechanics, CISM, 33100 Udine, Italy. Electronic Mail: 
%soldati@uniud.it.},
J.G.M. Kuerten$^2$,
B. Arcen$^3$, A. Tani\`{e}re$^3$,
G. Goldensoph$^4$, K.D. Squires$^4$,
M.F. Cargnelutti$^5$ and L.M. Portela$^5$
}
\address{
$^1$ Dept. Energy Technology and
Centro Interdipartimentale di Fluidodinamica e Idraulica,
University of Udine, Udine -- Italy\\
$^2$ Dept. Mechanical Engineering, Technische Universiteit
Eindhoven, Eindhoven -- Netherlands\\
$^3$ LEMTA, Nancy University, CNRS, ESSTIN, Vandoeuvre-l\'{e}s-Nancy -- France\\
$^4$ Dept. Mech. $\&$ Aerospace Engineering, Arizona
State University, Tempe -- USA\\
$^5$ Dept. Multi-Scale Physics, Technische Universiteit Delft,
Delft -- Netherlands
}
\maketitle

\newcommand{\mbf}{\mathbf}
\newcommand{\Pec}{\rm P_c}
\newcommand{\Pe}{\rm Pe}

\flushbottom
\topmargin  -16mm
\textwidth  140mm
\textheight 240mm
\columnsep   10mm
\parindent    4mm

%-------------------------------------------------------
%\makeatletter
%\global\@specialpagefalse
%\def\@oddhead{ \ifnum\c@page=1
%\let\@evenhead\@oddhead
%\llap{ \hfill Submitted to: {\em Int. J. Multiphase Flow} }
%\fi }
%\let\@evenfoot\@oddfoot
%\makeatother
%-------------------------------------------------------

\section*{\small\bf ABSTRACT}

In this paper, the results of an international collaborative
test case relative to the production of a Direct Numerical
Simulation and Lagrangian Particle Tracking database for
turbulent particle dispersion in channel flow at low Reynolds
number are presented. The objective of this test case is to
establish a homogeneous source of data relevant to the general
problem of particle dispersion in wall-bounded turbulence.
Different numerical approaches and computational codes have been
used to simulate the particle-laden flow and calculations
have been carried on long enough to achieve a
statistically-steady condition for particle distribution.
In such stationary regime,
a comprehensive database including both post-processed
statistics and raw data for the fluid and for the particles
has been obtained.
The complete datasets can be downloaded from the web at
{\sc{http://cfd.cineca.it/cfd/repository/}}.
In this paper, the most relevant velocity statistics
%of fluctuation intensities
%and Reynolds stresses
(for both phases) and particle distribution
statistics are discussed and benchmarked by
direct comparison between the different numerical
predictions.

%------------------------------------------------------
%\section*{\small\bf KEYWORDS}
%Direct Numerical Simulation, Lagrangian Tracking,
%Turbulent Boundary Layer, Coherent Structures,
%Deposition, Preferential Distribution, Re-entrainment.
%-------------------------------------------------------

%%%%%%%%%%%%%%%%%%%%%
\section{Introduction}
%%%%%%%%%%%%%%%%%%%%%

Turbulent particle dispersion in wall-bounded flows is a fundamental
issue in a number of industrial and environmental applications.
Direct Numerical Simulation (DNS) and Lagrangian Particle
Tracking (LPT) may be a useful tool to provide physical insights,
new modeling ideas and benchmark cases (Moin \& Mahesh, 1998;
Yeung, 2002). Despite the large number of published work, however,
it is extremely difficult to gather a uniform and complete source
of data that could be used to perform a phenomenological study of
some, still not well-established features of particle
transport in turbulent flows or to assess the effectiveness of
computer simulation models on the accuracy of predicted particle
deposition rates (Sergeev et al., 2002; Tian and Ahmadi, 2007).

Lack of uniformity and of completeness in the available
numerical data is connected to several reasons (associated with
the intrinsic complexity of turbulent transfer phenomena)
%which
%are correlated to a wide range of flow scales depending on the
%different interaction between particles and turbulent structures)
and is accompanied to uncertainty in methodologies, mostly due to the
large number of physical and computational parameters involved
and to the unclear influence of several of them.
The main physical parameters that will influence the simulation
results are the particle Stokes number, which quantifies
the response of the dispersed phase to the perturbations
produced by the underlying turbulence, and the flow Reynolds
number.
Other important parameters are related to
modeling of fluid-particle interaction
(one-way/two-way coupling); particle-particle interaction
(collision models); particle-wall interaction (reflecting
or absorbing wall, wall effects); particle rotation and
modeling of forces acting on particles (e.g.
the lift force).
On the computational side, the treatment of discrete particles
in DNS fields poses open or partly open questions on
the assessment of the performance of flow solvers that use
different numerical methods and on the accuracy of
the interpolation scheme used to obtain the fluid
velocity at the instantaneous particle location.
In this context, the proper choice of parameters such as
the grid resolution and the time step size required for
advancement of the governing balance equations becomes
extremely important.

%To investigate on these issues a systematic parametric study
%is required and reliable datasets are needed.
%
This paper is the result of the first necessary step
towards a rigorous, systematic analysis of these issues.
Specifically, the objectives of this analysis
are to have a large number of people working independently
on the same test case problem (DNS of particle dispersion
in turbulent channel flow) and to establish a large validated
database including
{\em (i)} reliable and accurate velocity statistics for the
fluid, for the particles and for the fluid at the
particle position (mean and rms velocities, skewness and
flatness, Reynolds stresses and quadrant analysis);
{\em (ii)} particle concentration profiles and deposition rates;
{\em (iii)} one-particle statistics (particle velocity
auto-correlations, particle turbulent diffusivity, particle
mean-square displacements, Lagrangian integral time scales);
{\em (iv)} two-particle statistics  (rms particle dispersion). 
Datasets come from five independent simulations and include
not only the post-processed statistics just listed
but also the corresponding raw data providing the evolution of the fluid
velocity field and the time behavior of the particle position
and velocity components: these data are made available to
users who need to compute specific statistics other than those
included in the database.
%In addition to these standard statistics, raw
%data providing the evolution of the fluid velocity
%field and the time behavior of the particle position
%and velocity components will also be included for further
%post-processing analyses.
Besides providing a homogeneous source of data
on DNS and LPT not previously available, the database
can be used as benchmark either to compare
directly different numerical approaches or
to validate engineering models for particle dispersion
(e.g. two-fluid Eulerian models).
The need for this type of data could be extended
also to commercial softwares for computational fluid
dynamics: these softwares, even though usually exploited for
high-Reynolds-number flows in complex geometries, fail predictions
of multiphase flows due to the lack of appropriate physical models
for particle dispersion, resuspension and deposition.

The test case was conceived in 2004 at the IUTAM Symposium
on Computational Approaches to Multiphase Flow
(Balachandar and Prosperetti, 2006) and
%\cite{iutam} and
%(October 04 - 07, 2004), organized by
%Prof. S. Balachandar and Prof. A. Prosperetti, and
%held at the Argonne National Laboratory (Illinois,
%USA).
it was first advertised in 2005 at the $11^{th}$ Workshop on
Two-Phase Flow Predictions (Sommerfeld, 2005).
%\cite{merseburg}.
%(April 5 - 8, 2005),
%organized by Prof. M. Sommerfeld, and held at the
%Universit\"{a}t Halle-Wittenberg (Merseburg,
%Germany).
During the workshop, common base guidelines
for participant groups were provided.
The following groups (listed in random order)
joined the test case calculations:
1) C. Marchioli and A. Soldati (Group UUD hereinafter),
2) J.G.M. Kuerten (Group TUE hereinafter),
3) B. Arcen and A. Tani\`{e}re (Group HPU hereinafter),
4) G. Goldensoph and K. Squires (Group ASU hereinafter),
5) M.F. Cargnelutti and L.M. Portela (Group TUD hereinafter).
As starting point of the test case, a DNS
of dilute particle-laden turbulent channel flow at low
Reynolds number has been performed by all groups following
the base guidelines.
Aim of this benchmark calculation is to build a thorough
statistical framework including both statistically-developing
and statistically-steady conditions
for the distribution of the dispersed phase.
%and, to eventually single out the effect of each
%simulation parameter from macroscopic particle
%behavior.
To quantify the collaborative effort required
by the test case, it should be
noted that the simulation time taken
for each group to achieve a statistically-steady
condition for the particle distribution was of
the order of eight to ten months, mostly depending
on the availability of computational resources.
%depending on the different code used and
%on the different production machine available. 
This is equivalent to an overall simulation
time of about four years on standard production machines.

The present paper
%reports on the results produced by the
%benchmark calculation, that constitutes the starting point
%of the test case, and
is organized as follows:
first the physical problem and the
numerical methodology adopted by each group
are briefly outlined, then the performance of the
different numerical approaches is benchmarked through
direct comparison of the most relevant
statistics for both phases are discussed. In the final section,
conclusions and implications for future developments
of the test case are drawn.

%%%%%%%%%%%%%%%%%%%%%
\section{Physical Problem and Numerical Methodology}
%%%%%%%%%%%%%%%%%%%%%

\subsection{Particle-laden turbulent channel flow}

The flow into which particles are introduced is a turbulent
channel flow of gas.
In the present study, we consider air (assumed
to be incompressible and Newtonian)
with density $\rho = 1.3~kg~m^{-3}$ and kinematic
viscosity $\nu = 15.7{\times}10^{-6}~m^{2}~s^{-1}$.
The governing balance equations for the fluid (in
dimensionless form) read as:
\begin{equation}
\label{cont}
\frac{\partial{u_i}}{\partial{x_i}} = 0,
\end{equation}
\begin{equation}
\label{ns}
\frac{\partial{u_i}}{\partial{t}} = -u_{j}\frac{\partial{u_i}}{\partial{x_j}} +
\frac{1}{Re}\frac{\partial^{2}{u_i}}{\partial{x_j}^2} - \frac{\partial{p}}{\partial{x_i}}
+ \delta_{1,i},
\end{equation}
where $u_i$ is the $i^{th}$ component of the dimensionless velocity
vector, $p$ is the fluctuating kinematic pressure, $\delta_{1,i}$
is the mean dimensionless pressure gradient that drives the flow
and $Re_{\tau} = u_{\tau}h/\nu$ is the shear Reynolds number
based on the shear (or friction) velocity, $u_{\tau}$,
and on the half channel height, $h$.
%and on the fluid kinematic viscosity, $\nu$.
The shear velocity is defined as $u_{\tau} = (\tau_{w}/\rho)^{1/2}$,
where $\tau_{w}$ is the mean shear stress at the wall.
%and $\rho$ is the fluid density.
In this benchmark calculation, the shear Reynolds number is $Re_{\tau} = 150$;
the corresponding bulk Reynolds number is $Re_{b} = u_{b}h/\nu = 2100$ based on
the bulk velocity $u_b=1.65~m~s^{-1}$.
%and on $h$.
All variables considered in this study are reported in dimensionless
form, represented by the superscript +\footnote{The superscript + has been dropped from Eqns.
(\ref{cont}) and (\ref{ns}) for ease of reading.} and expressed in wall units. Wall units are
obtained combining $u_{\tau}$, $\nu$ and $\rho$.
%
% and the half channel height, $h$.
%

\begin{figure}
\centerline{\includegraphics[height=8.0cm]{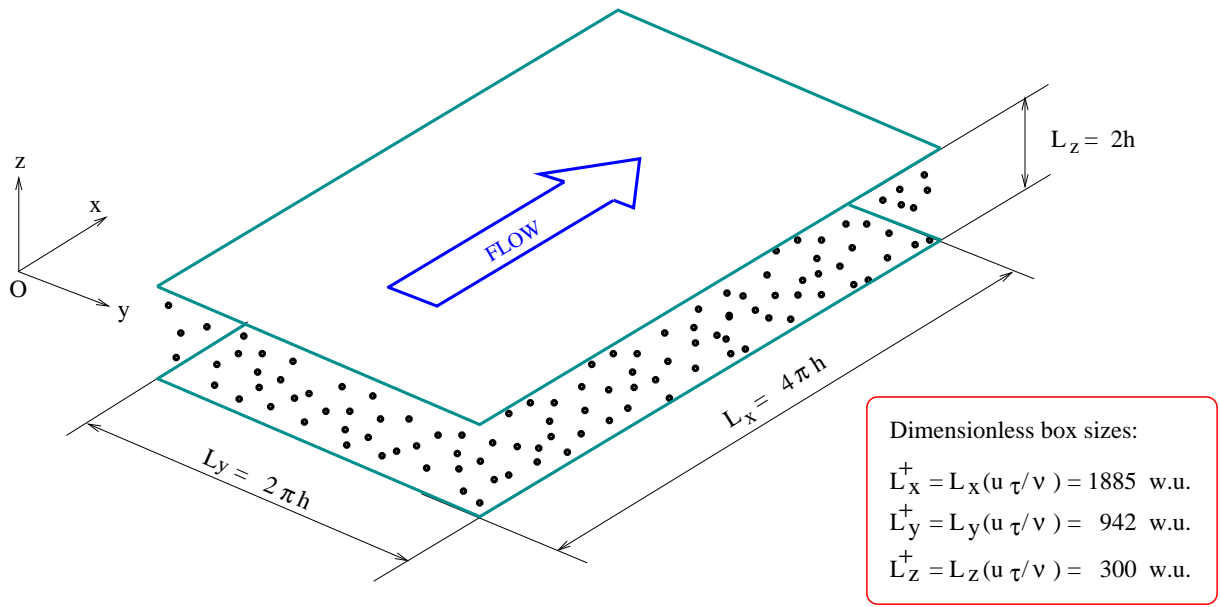}}
\caption{Particle-laden turbulent gas flow in a flat channel: computational domain.}
\label{f-chandomain}
\end{figure}

The reference geometry consists of two infinite
flat parallel walls: the origin of the coordinate system is located
at the center of the channel and the $x-$, $y-$ and $z-$ axes point
in the streamwise, spanwise and wall-normal directions respectively
(see Fig.~\ref{f-chandomain}).
Periodic boundary conditions are imposed on the fluid velocity field
in $x$ and $y$, no-slip boundary conditions
are imposed at the walls.
The calculations were performed
on a computational domain of size
$4 \pi h \times 2 \pi h \times 2 h$,
corresponding to $1885\times942\times300$ wall
units in $x$, $y$ and $z$ respectively.
%
%discretized with $128\times128\times129$ nodes.
%
For ease of reading, details on the Eulerian grid used
to discretize the flow domain and on the time step
size, $\Delta t^+$, employed by each group are given in
Section 2.2. Here, we just mention that the base
simulation requirements prescribe a minimum number
of grid points in each direction to ensure that the
grid spacing is always smaller than the smallest
flow scale \footnote{In the present flow configuration,
the non-dimensional Kolmogorov length scale, $\eta_K^+$,
varies along the wall-normal direction from a minimum
value $\eta_K^+=1.6$ at the wall to a maximum value
$\eta_K^+=3.6$ at the centerline.
In terms of time scales, the Kolmogorov time
scale, $\tau_K^+$, varies along
the wall-normal direction from a minimum
value $\tau_K^+=2$ at the wall to a maximum value
$\tau_K^+=13$ at the centerline (Marchioli et al., 2006).}
and that the limitations imposed
by the point-particle approach are satisfied.
%%\ref{meth}.
%
%${\Delta}t^{+} = 0.045$ in wall time units.
%

Particles with density $\rho_p=1000~kg~m^{-3}$ are injected
into the flow at concentration low enough to
consider dilute system conditions (particle-particle interactions
are neglected).
Furthermore, particles are
assumed to be pointwise, rigid and spherical. The motion of
particles is described by a set of ordinary differential
equations for particle velocity and position at each time step.
For particles much heavier than the fluid ($\rho_{p}/\rho \gg 1$)
Elghobashi and Truesdell (1992) have shown that the only
significant forces are Stokes drag and buoyancy
and that Basset force can be neglected being
an order of magnitude smaller.
In the base simulation, the aim is to minimize the
number of degrees of freedom by keeping the simulation setting
as simplified as possible; thus
the effect of gravity has also been
neglected. With the above assumptions the following
Lagrangian equation for the particle velocity is obtained:
\begin{equation}
\label{part}
\frac{d{\bf u}_p}{dt} =
-\frac{3}{4}\frac{C_D}{d_p}\left( \frac{\rho}{\rho_p} \right) |{\bf u}_p - {\bf u}|({\bf u}_p - {\bf u}),
\end{equation}
where ${\bf u}_p$ and ${\bf u}$ are the particle
and fluid velocity vectors, $d_p$ is the particle diameter and
$C_D$ is the drag coefficient given by (Rowe and Enwood, 1962):
\begin{equation}
C_D = \frac{24}{Re_p}(1+0.15Re_{p}^{0.687}),
\end{equation}
where $Re_p$ is the particle Reynolds number ($Re_p = d_{p}|{\bf u}_{p} - {\bf u}|/\nu$).
The correction for $C_D$ is necessary because $Re_p$ does
not necessarily remain
small, in particular for depositing particles.

For the simulations presented here,
three particle sets were considered, characterized by
different relaxation times, defined as
$\tau_{p} = \rho_{p}d^{2}_{p}/18\mu$
where $\mu$ is the fluid dynamic viscosity.
The particle relaxation time is made
dimensionless using wall variables
and the Stokes number for each particle
set is obtained as $St=\tau_p^+=\tau_p/\tau_f$
where $\tau_f=\nu/u_{\tau}^2$ is the
characteristic time scale of the flow.
Table~\ref{part} shows all the parameters
of the particles injected into the flow field.
To build the database, the ``one-way coupling'' approximation
(under which particles do not feedback on the flow field)
was considered.
At the beginning of the Lagrangian tracking, particles
were distributed randomly over
the computational domain and their initial
velocity was set equal to that of the fluid
at the particle initial position.
%Periodic boundary conditions are imposed on particles
We remark here that the process of particle dispersion will
not be sensitive to this initial condition if the
long-term features of the motion are investigated.
%in both streamwise and spanwise directions,
%elastic reflection is applied when
%the particle center is less than a distance
%$d_{p}/2$ from the wall.
Regarding the boundary conditions of the dispersed phase,
perfectly-elastic collisions at the smooth wall were assumed
when the particle center was at a distance from the wall
lower than one particle radius. Further, particles moving
outside of the computational domain in the streamwise and/or
spanwise directions were reintroduced via periodicity.

Further details on the Lagrangian tracking (e.g. the numerosity
of particle sets, the fluid velocity interpolation scheme, etc.)
are given for each group in Section 2.2.

\subsection{DNS methodology and computational resources}
\label{meth}

In this Section, the different numerical approaches and computational
codes are briefly outlined.
They are also summarized in Table \ref{dns}, where
the {\em numerical} degrees of freedom characterizing the benchmark
calculation are presented. 
The possibility of using different numerical schemes and/or
different values for some simulation parameters (like the time
integration step size or the number of grid points, for instance)
allows clearcut evaluation of how the accuracy of the DNS results
depends on the choice made.
%A reference is also provided, where the reader
%can find further details.

\begin{itemize}
\item Group UUD:
the computational flow solver is based on the Fourier-Galerkin
method in the streamwise and spanwise directions, whereas a
Chebyshev-collocation method in the wall-normal direction.
Time integration of fluid uses a $2^{nd}$-order Adams-Bashforth
scheme for the non-linear terms (which are calculated in a
pseudo-spectral way with de-aliasing in the periodic directions)
and an implicit Crank-Nicolson scheme for the viscous terms.
A LPT code coupled with the DNS code
is used to calculate particles paths in the flow field.
The particle equation of motion is solved using a
4th order Runge-Kutta scheme for time integration.
Fluid velocities at particle position are obtained
using 6th-order Lagrangian polynomials: near the
wall, the interpolation scheme switches to one-sided.
The total number of particles tracked is $10^5$.
The computational time step size in wall units is
${\Delta}t^+ = 0.045$ for the fluid
%${\Delta}t^{+} = 0.045$ in wall time units.
and ${\Delta}t^+ = 0.45$ for the particles,
this latter value being larger than those
adopted by the other groups (see Table II).
Simulations were performed running a serial version of the code
on a standard production machine with Pentium IV 2.6GHz CPU and
1Gb RAM.
Further details about the numerical methodology of this group
can be found in Marchioli and Soldati (2002).
\item Group TUE:
the computational flow solver is based on the Fourier-Galerkin
method in the streamwise and spanwise directions, whereas a
Chebyshev-collocation method is used in the wall-normal direction.
Non-linear terms are calculated in a pseudo-spectral way with
de-aliasing in the periodic directions.
The solution is completely divergence-free through the use of
the influence matrix method with full correction
to remove Chebyshev-truncation errors.
Time integration is performed with a 3rd-order three-stage
Runge-Kutta method for the non-linear terms and the implicit
Crank-Nicolson method for the linear terms.
The particle equation of motion is solved with the second-order
Heun method. 
Fluid velocity at particle position is obtained by fourth-order
accurate interpolation: Lagrange polynomials are used in the
periodic directions whereas Hermite polynomials are used in
the wall-normal direction.
%Particles bounce elastically against the walls.
The total number of particles tracked
is $10^5$.
The computational time step in wall units
is ${\Delta}t^+ = 0.032$ for both phases.
Simulations were performed running a mpi-parallelized version of the code
on a Linux PC cluster with 8 CPU and 8 GB RAM.
Further details about the numerical methodology of this group
can be found in Kuerten (2006).
\item Group HPU:
a 2nd-order finite-difference DNS solver,
based on the model proposed by Orlandi (2000), was
used for the flow: time discretization is semi-implicit,
i.e. the
non-linear terms are written explicitly with a
third-order Runge-Kutta scheme and the viscous
terms are written implicitly using a Crank-Nicolson
scheme.
%
%In the wall-normal direction, the mesh is
%stretched according to a hyperbolic tangent law,
%whereas a uniform mesh is applied in the streamwise
%and spanwise directions.
%
%Note that the size of the flow domain in the streamwise,
%wall-normal, and spanwise direction is
%$4 \pi \delta \times 2 \delta \times 2 \pi \delta$
%discretized on a $192 \times 128 \times 160$ grid.
%
Computations were run with imposed flow
rate corresponding to a bulk Reynolds
number $Re_{b} = 2280$ based on
the bulk velocity and the channel half-width.
The shear Reynolds number obtained at steady state
is $Re_{\tau} = 155$, slightly higher than that
simulated by the other groups.
To initialize position and velocity of the particle
phase, the flow domain was divided into 128 slices along
the wall-normal direction, the thickness of each
slice being equal to the wall-normal grid spacing.
Samples of 5000 solid particles were distributed
homogeneously within each slice.
The total number of particles tracked
is thus $6.4 \cdot 10^5$.
The computational time step in wall units
is $t^+ = 0.05$ for both phases.
Simulations were performed running a serial version of the code
on a standard production machine with Pentium IV 3.4Ghz CPU and 3Gb RAM.
Further details about the numerical methodology of this group
can be found in Arcen et al. (2006).
\item Group ASU:
a fractional-step method is used to solve for the NS equations.
Spatial derivatives are evaluated using $2^{nd}$-order centered
differences.
Time integration of fluid is performed using a $2^{nd}$-order
Adams-Bashforth scheme for the non-linear terms and an implicit
Crank-Nicolson scheme for the viscous terms.
A $2^{nd}$-order Adams-Bashforth scheme is also used for time
integration of particle equation of motion, and $3^{rd}$-order
Lagrange polynomials are used for fluid velocity interpolation.
The total number of particles tracked
is $10^5$.
The computational time step in wall units
is $t^+ = 0.05$ for both phases.
Simulations were performed running a serial version of the code
on a Pentium IV 2.6GHz CPU and 1Gb RAM.
Further details about the numerical methodology of this group
can be found in Goldensoph (2006).
\item Group TUD:
a standard fine-volume code based on a predictor-corrector solver
is used to solve for the NS equations on a staggered grid.
Time integration of fluid is performed using a $2^{nd}$-order
Adams-Bashforth scheme.
A $2^{nd}$-order Runge-Kutta scheme is also used for time
integration of particle equation of motion, and tri-linear
interpolation is used to calculate the fluid velocity at the
particle position.
The total number of particles tracked
is around $9.5 \cdot 10^5$.
The computational time step for the
fluid is roughly $t^+ = 0.026$ in wall units,
the exact size being determined adaptively.
The time step size for the particles is
always equal to or smaller than the fluid
time step since it is chosen such that a
particle can not travel more that half grid
cell per iteration.
Simulations were performed running a serial
version of the code, partially on a standard
production machine with AMD Athlon 2133 MHz dual processor CPU and 2 GB RAM, and
partially on one node of a SGI Altix 3700 system consisting of 416 CPUs
(Intel Itanium 2, 1.3 GHz each) and 832 GB total RAM.
Further details about the numerical methodology of this group
can be found in Portela and Oliemans (2003).

\end{itemize}
 
%%%%%%%%%%%%%%%%%%%%%
\section{Results}
%%%%%%%%%%%%%%%%%%%%%

In this Section some of the most relevant statistics
for the fluid phase and for the particle phase are
presented and discussed {\em vis-\a-vis} to benchmark the
performance of the different numerical approaches.
It is important to remind the reader that all particle statistics
shown in this paper refer to a steady state for particle
distribution.
Particle statistics were computed summing
the desired variable (velocity, velocity fluctuation, etc.)
over all particles in a certain sampling volume, constituted by
wall-parallel fluid slab obtained as described in Sec. 3.2,
%\ref{partstat})
and averaging by the number of particles in the sampling volume.
This type of density-weighted statistics is particularly useful
for model developers.

For sake of brevity, we will limit our analysis
to the first- and second-order moments of both phases
(namely the mean streamwise velocity and the rms values
of the three velocity components), to the Reynolds stresses
and to the particle concentration profiles.
Higher order statistics
as well as two-particle statistics will not be presented here
since they would not add to the discussion. The reader is referred
to the raw data repository for further statistical exploration.

\subsection{Fluid statistics}

\begin{figure}
\centerline{
\includegraphics[width=10.0cm]{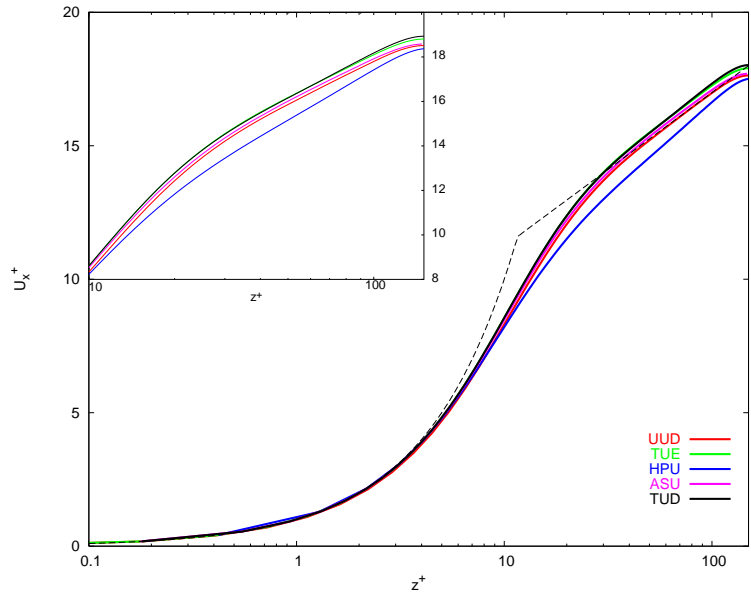}
}
\vspace{0.5cm}
\caption{Mean streamwise fluid velocity.}
\label{mean-fluid}
\end{figure}

Fig. \ref{mean-fluid} shows the mean streamwise fluid velocity
profiles, $U_x^+$. The profile of each group is represented by a
solid colored line
%({\bf{Q: should we use symbols instead of colored lines?}})
whereas
the black solid line represents the mean velocity profile given by the
law of the wall, $U_x^+=z^+$, and by the log law, $U_x^+=2.5 \log z^+ + 5.5$.
It is apparent that the profiles almost overlap, yet a close-up view has been
included in the diagram to appreciate better
the behavior of $U_x^+$ in the outer layer of the channel,
where a slight velocity deficit is observed for the profile of Group
HPU.
This may be due to the smaller number of grid points taken to
discretize the domain along the wall-normal direction ($N_z=128$)
compared to those taken by Group TUD ($N_z=192$) using the same
second-order accurate flow solver.
%which uses a finite-difference flow solver with imposed flowrate.

\begin{figure}
\centerline{
\includegraphics[width=10.0cm]{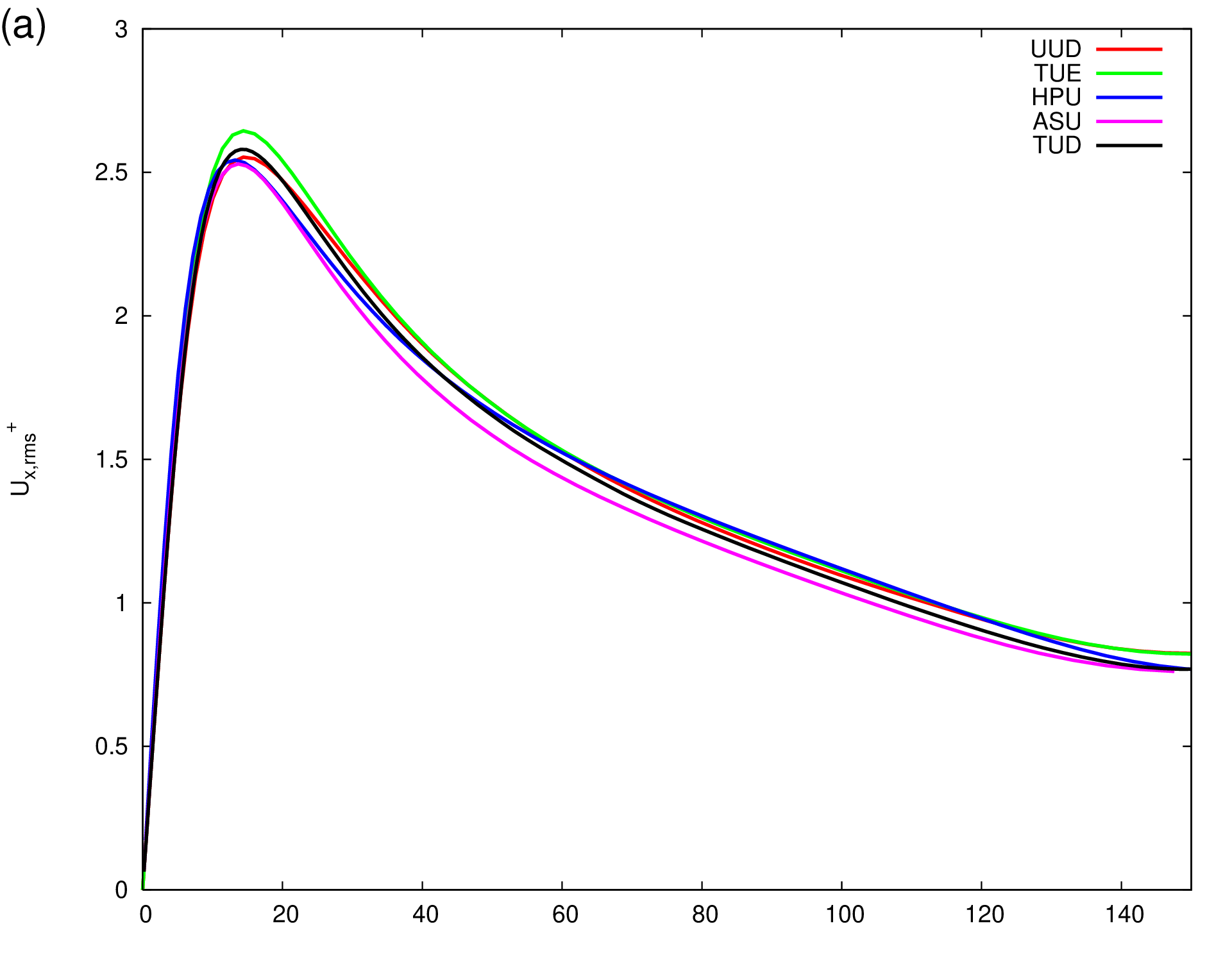}
}
\vspace{-0.5cm}
\centerline{
\includegraphics[width=10.0cm]{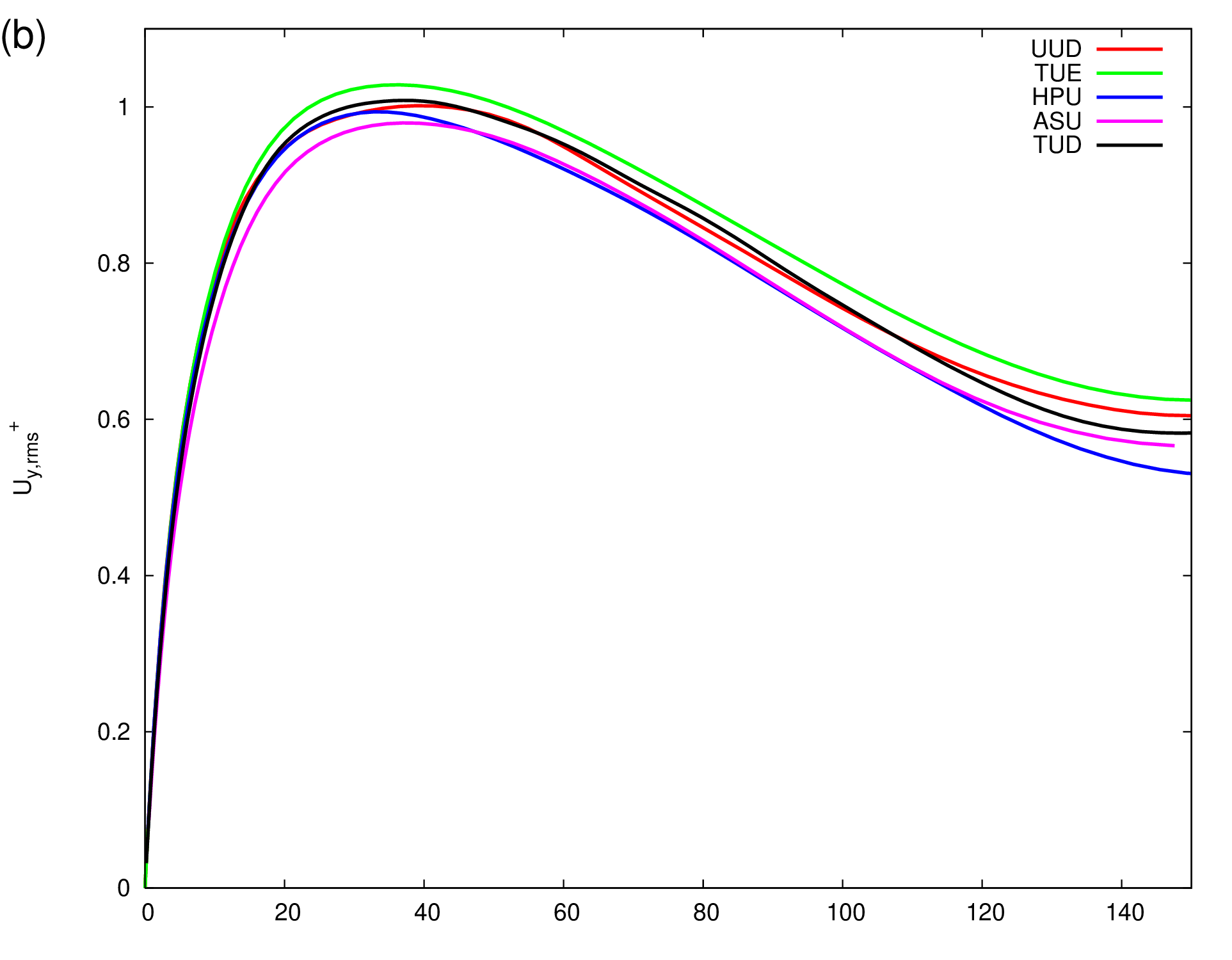}
}
\vspace{-0.5cm}
\centerline{
\includegraphics[width=10.0cm]{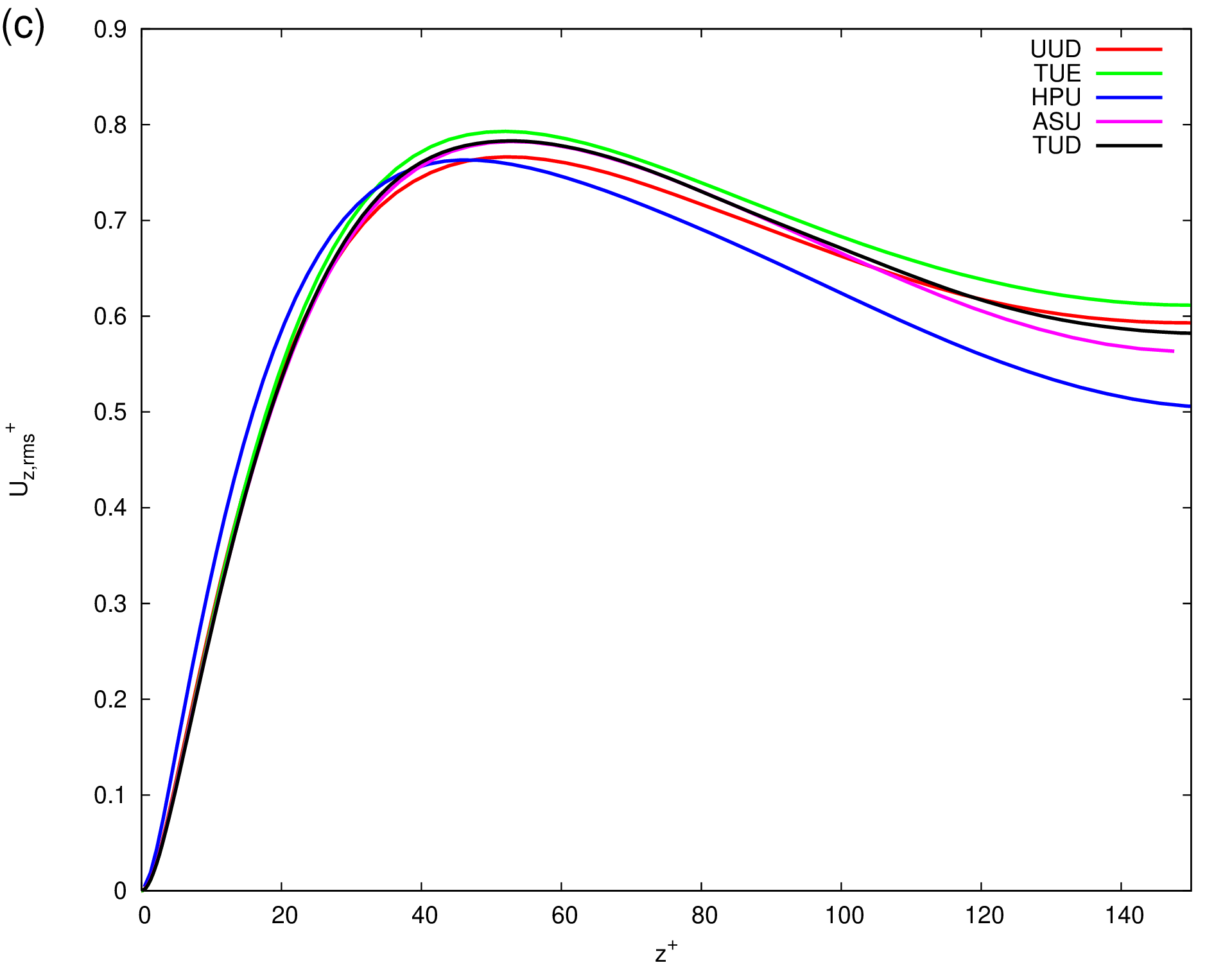}
}
%\vspace{0.5cm}
\caption{Root mean square of fluid velocity fluctuations, $U_{i,rms}'$.}
\label{rms-fluid}
\end{figure}

Figs. \ref{rms-fluid}a-c show the root mean square (rms) of the fluid velocity
($U_{i,rms}'$) in the streamwise ($i=x$), spanwise ($i=y$) and wall-normal
($i=z$) direction, respectively. The color code is the same as before.
Results are in rather good agreement for the $U_{x,rms}'$ component but
small differences in the quantitative numbers arise for 
$U_{y,rms}'$ and $U_{z,rms}'$, particularly outside the buffer
layer close to the channel centerline.
Differences observed in the first and second order moments of the
fluctuating fluid velocity field are, of course, due solely to the
specific numerical method employed by each flow solver
and to the accuracy of grid discretization. These differences in
modeling the flow field will add to differences in modeling
the particle motion and will show up also in the statistical
moments for the particle velocity.

\begin{figure}
\centerline{
\includegraphics[width=10.0cm]{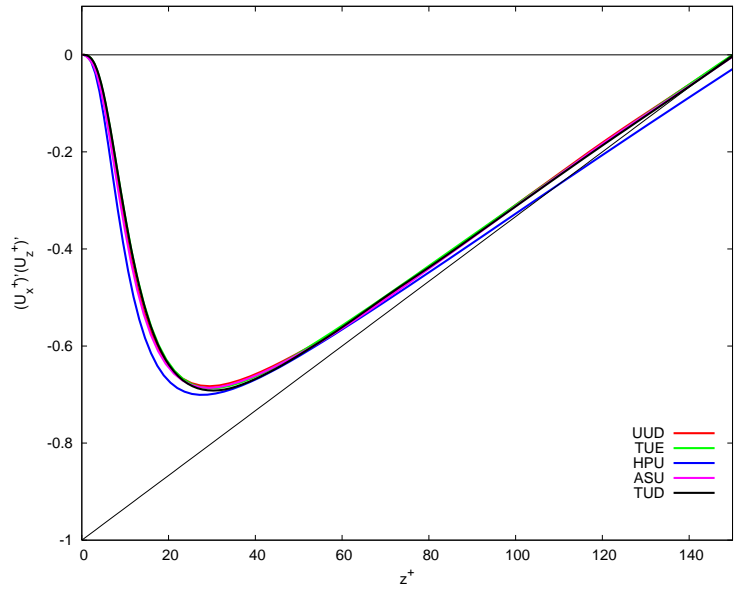}
}
\caption{Fluid Reynolds stress: ($U_x^+)'(U_z^+)'$ component.
}
\label{strflu}
\end{figure}

To conclude the section devoted to fluid statistics, the results
obtained for the Reynolds stresses are discussed since they may
be of interest in the context of deriving models for fully-developed,
dilute particulate turbulent flow with a suitably closed system of
governing equations (Sergeev et al., 2002). Specifically,
Fig. \ref{strflu} shows the time- and space-averaged $(U_x^+)'(U_z^+)'$-component
of the Reynolds stress tensor, $(U_x^+)'$ and $(U_z^+)'$ being the
non-dimensional fluid velocity fluctuations in the streamwise direction and in the
wall-normal direction, respectively. The agreement between the different
profiles is indeed satisfactory, particularly if one considers that
the profile of Group HPU, which does not overlap perfectly around the
negative peak value and near the centerline, was obtained for a slightly
higher value of the shear Reynolds number and thus goes to zero at
$z^+=155$ rather than $z^+=150$.

\subsection{Particle statistics}
\label{partstat}

When computing particle statistics, it is of particular importance
to define precisely the computational procedure to
ensure reproducibility of the results. In this study, particle
statistics were computed
by averaging over $N_s=193$ wall-parallel fluid slabs distributed
non-uniformly along the wall-normal direction. The thickness
of the $s^{th}$ slab, $\Delta z^+(s)$, was obtained by means of hyperbolic-tangent
binning with stretching factor $\gamma=1.7$:
\begin{equation}
\Delta z^+ (s) = \frac{Re_{\tau}}{tanh(\gamma)}
\left[
tanh \left( \gamma \frac{s}{N_s} \right)-
tanh \left( \gamma \frac{s-1}{N_s} \right)
\right]
\end{equation}
The smallest
thickness is at the wall ($\Delta z^+_{min}=0.361$)
whereas the largest thickness is at
the channel centerline ($\Delta z^+_{max}=2.84$).
Despite the large particle
concentration gradients expected near the walls,
$\Delta z^+_{min}$ was chosen slightly larger than the wall-normal
grid spacing of the numerical simulation, the minimum
thickness allowed at the wall being limited by the $St=25$
particle radius (see Table \ref{part}).
A particle belongs to a slab if its center
is located inside the slab.

\begin{figure}[hb]
\centerline{
\includegraphics[height=9.0cm]{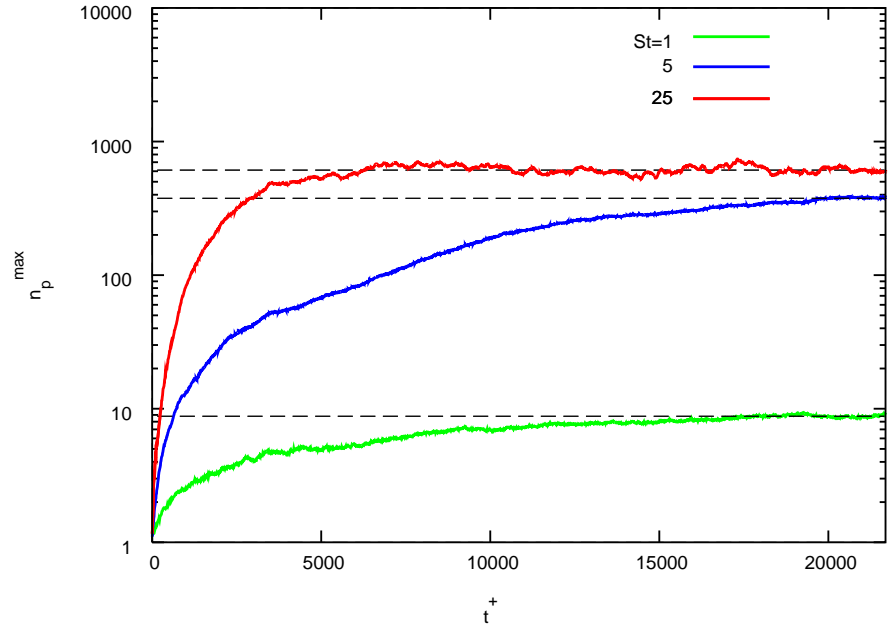}
}
\vspace{0.5cm}
\caption{Maximum value of particle number density at the wall,
$n_{p}^{max}$, as function of time $t^{+}$ (lin-log plot).
Data from Group UUD.}
\label{conctime}
\end{figure}

Since the aim of the benchmark simulation is to reach
a statistically-steady state for the particle distribution,
the process of accumulation was followed over time
starting from an initial condition of randomly-distributed
particles.
Fig.~\ref{conctime} shows the time evolution of
the maximum value of particle number density near the wall,
$n_p^{max}$, for each particle set as obtained by the Group
UUD (profiles obtained by the other groups are not shown
as they provide qualitatively similar results and would
not add to the discussion)
up to $t^+=21150$.
The rationale for monitoring
this quantity lies in the
fact that the concentration close to the wall is
the one that takes longer to reach its
steady state.
After an initial large change covering a time span
of about 1000 wall time units and a slow asymptotic
convergence towards a {\em mean} value (represented by
the horizontal dashed lines in Fig.~\ref{conctime}),
this state appears to be achieved at $t^+ \simeq 20000$.
%As mentioned by Portela et al. (2002), it takes an enormous
%amount of time to reach a statistically-steady particle
%concentration. In the present benchmark simulation,
%a non-dimensional time $t^+ = 20000$ in wall units
%was required.
Since the non-dimensional bulk
velocity in the channel is roughly equal to $U_x^+=15$,
this time threshold corresponds to a {\em developing-length}
of roughly 1000 channel heights. This number shows the difficulties,
both numerical and experimental, of obtaining information
on fully developed particle-laden channel flows:
the {\em particle developing-length} can be much larger
than the hydrodynamic developing-length, requiring extremely
long computational times in numerical simulations, and
extremely long channels (or pipes) in physical experiments.

\begin{figure}[h]
\vspace{-1.3cm}
\begin{center}
\includegraphics[width=7.0cm,angle=-90]{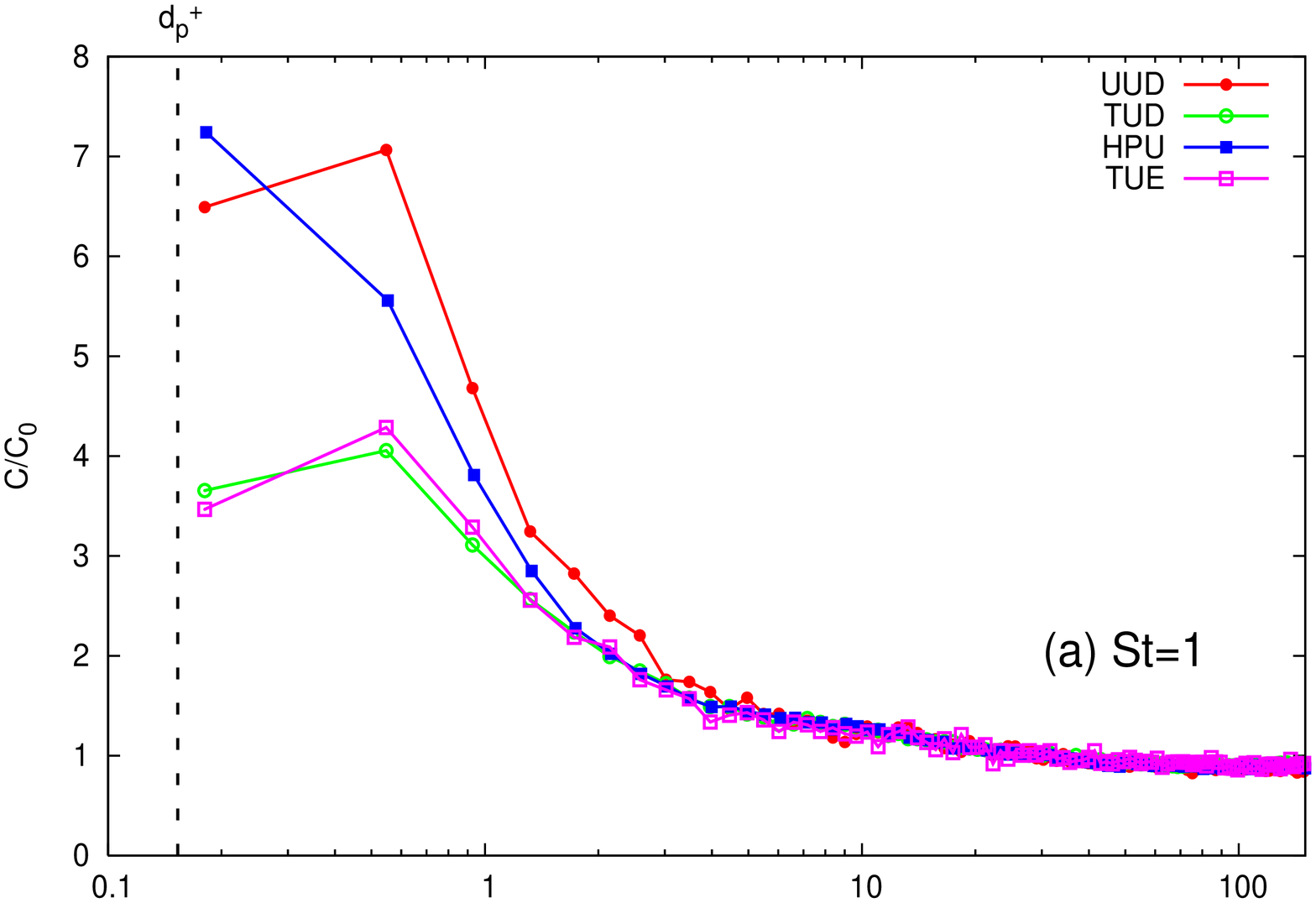}
\end{center}
\vspace{-1.4cm}
\begin{center}
\includegraphics[width=7.0cm,angle=-90]{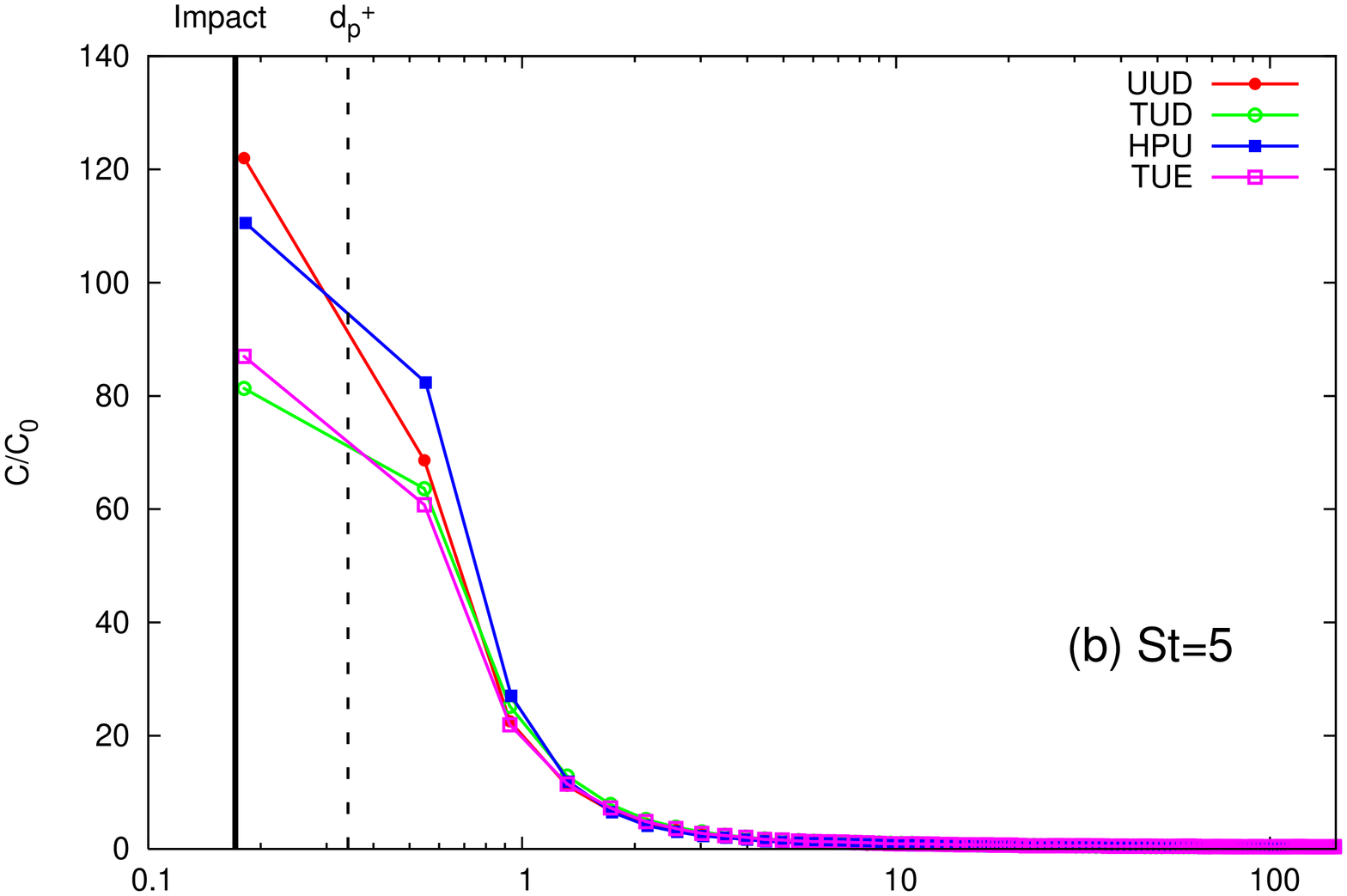}
\end{center}
\vspace{-1.4cm}
\begin{center}
\includegraphics[width=7.0cm,angle=-90]{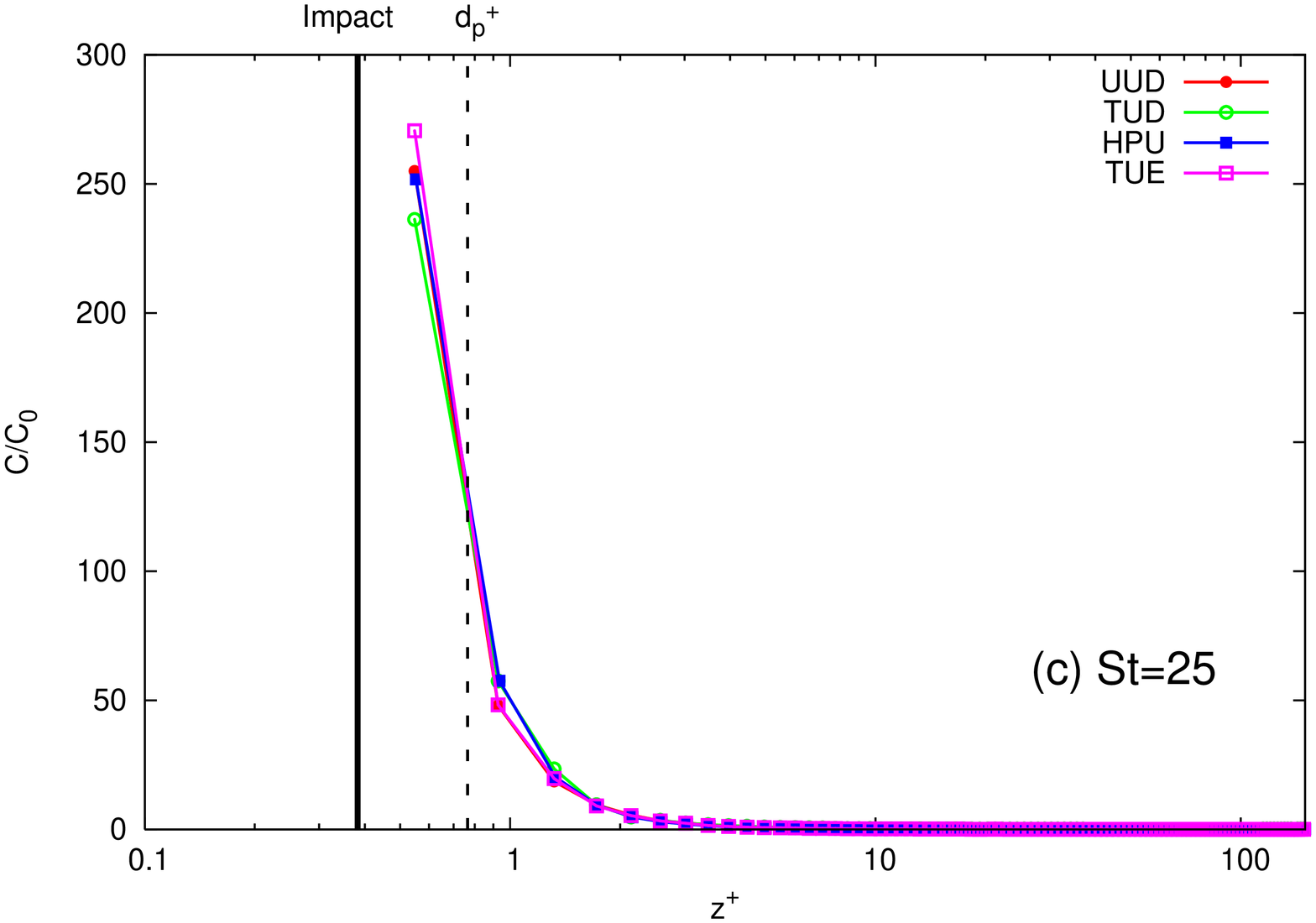}
\end{center}
\caption{Instantaneous particle number density
wall-normal profiles (log-log plot): (a) $St = 1$,
(b) $St=5$, (c) $St=25$. The vertical solid line in each
diagram indicates the position where the particles hit the wall
({\em Impact}); the vertical dash-dotted line gives a visual indication
of the size of the particle diameter in wall units ($d_p^+$).}
\label{conc}
\end{figure}

Fig.~\ref{conc} shows the particle concentration
profiles, $C/C_0$, as function of the wall-normal
coordinate, $z^{+}$, at
%$t^{+} \approx 1000$ and
the threshold $t^{+} = 20000$ and for each particle set.
Concentration statistics at earlier stages
of the simulations, in which a statistically-developing
condition for the particle concentration exists, have also been
computed:
datasets have been gathered past the initial transient of
1000 wall time units, taken to obtain
results that are independent of 
the initial conditions imposed on the particles, up
to the end of the simulations with saving frequency of 1000 wall time units.

Particle concentration has been obtained as follows:
first, the flow domain is divided into slabs according
to the above-mentioned binning procedure;
second, at each time step the number of particles
within each slab is determined and divided by the
volume of that slab to obtain the local
concentration $C=C(s)$; finally, $C$ is normalized by
its initial value, $C_0$.
According to this procedure, the ratio $C/C_0$ is in fact a
particle number density distribution and will be larger than unity in the flow regions were
particles tend to preferentially distribute and smaller than unity in the regions
depleted of particles.
%
%A log-log scale is used to capture the different
%behavior of the profiles in the proximity
%of the wall and to underline the
%different magnitude of particle number density for
%each particle set.
%In the range of particle Stokes numbers
%considered, we observe a build-up of the particle
%number density in the wall region, where the
%statistically-steady concentration reaches very
%high values.
%

From Fig.~\ref{conc} we observe that, starting from
an initial distribution corresponding to a
flat profile centered around $C/C_0 \simeq 1$,
the expected near-wall concentration build up occurs
and that the magnitude of this
build up depends on the Stokes number of the particles.
The steady-state concentration profiles,
albeit being not physical (near-wall effects arise
that are more complicated than those taken into account
in the base simulation), are well consistent with
all the information a DNS can provide.
Also, the agreement between the profiles of each group
increases with the particle Stokes number.
For the two smaller particle sets ($St=1$
%\footnote{{\bf Note:
%the steady-state concentration for the $St=1$ particles
%from Group ASU is not available.}}
and $St=5$),
quantitative differences in the predicted
near-wall peak values are observed among all groups;
however, the results of Group TUE match
closely with those of Group TUD and there is a good
agreement between the results
of Group UUD and those of Group HPU (even if the shape
of the profiles and the location of the peak value is
different).
%whereas the results of Group TUE match
%closely with those of Group TUD.
%Overall, quantitative differences in the predicted
%near-wall peak values are observed among all groups
Differences are still present for the higher
inertia particles ($St=25$), yet they become less
evident in proportion.
Unfortunately, for the $St=1$ particles, it was not
possible to include data from Group ASU due to their
unavailability.
Discrepancies in the quantification of local
particle concentration
arise not only because of the diverse
numerical schemes and grid discretizations adopted by each group but
also, if not mainly, because of the numerical errors associated
{\em (i)} with the different interpolation techniques used to obtain the
fluid velocity at particle position and {\em (ii)} with the choice
of the time step size used to integrate the
equation of motion for the particles.
These numerical errors sum up over time and give the accumulated profile
deviations observed in Fig.~\ref{conc}.
%On top of this, the concentration profiles shown in Fig.~\ref{conc}
%are instantaneous and further deviations due to fluctuations
%are thus expected.
It should be noted, however, that concentration profiles start to
deviate significantly from each other only very close to the wall
(roughly within one wall unit from the wall, a thickness that could be
considered negligible from a pragmatic engineering perspective)
and that deviations are magnified by the log-lin scale chosen to
visualize the profiles.
Also, we remark that the presence of discrepancies
does not imply that only one of the profiles shown is correct
while the others are wrong; rather, it implies that there might
be a {\em best} prediction for a given statistical
quantity which, however, is not known a priori.
In other words, with the current data available it is not
possible to conclude which is, if any, the best dataset.
However, we can observe that the range of wall concentration
predictions can be accepted as a good measure of particle
wall concentration under the modelling assumptions used
in this work.
%Elaborating on this observation, it seems very difficult to provide clearcut
%indications on the reliability of each dataset or to indicate
%which prediction is more accurate in absolute terms.

%\begin{itemize}
%\item fluid velocity statistics (mean and rms, Reynolds stresses)
%\item particle velocity statistics (mean and rms, Reynolds stresses)
%\item particle concentration profiles
%\item other (one-particle/two-particle statistics)?
%\end{itemize}

\begin{figure}
\centerline{
\includegraphics[width=10.0cm]{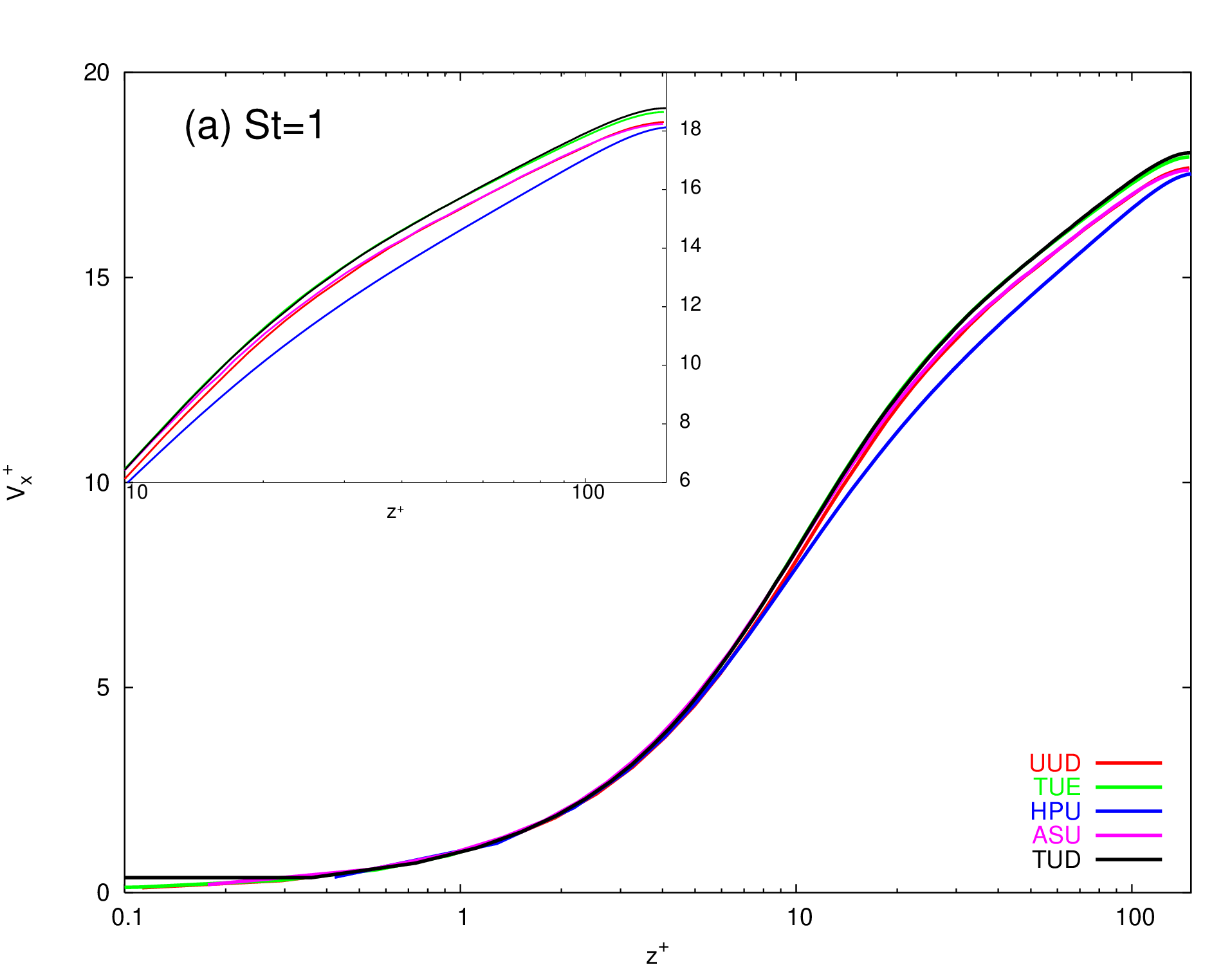}
}
\vspace{-0.5cm}
\centerline{
\includegraphics[width=10.0cm]{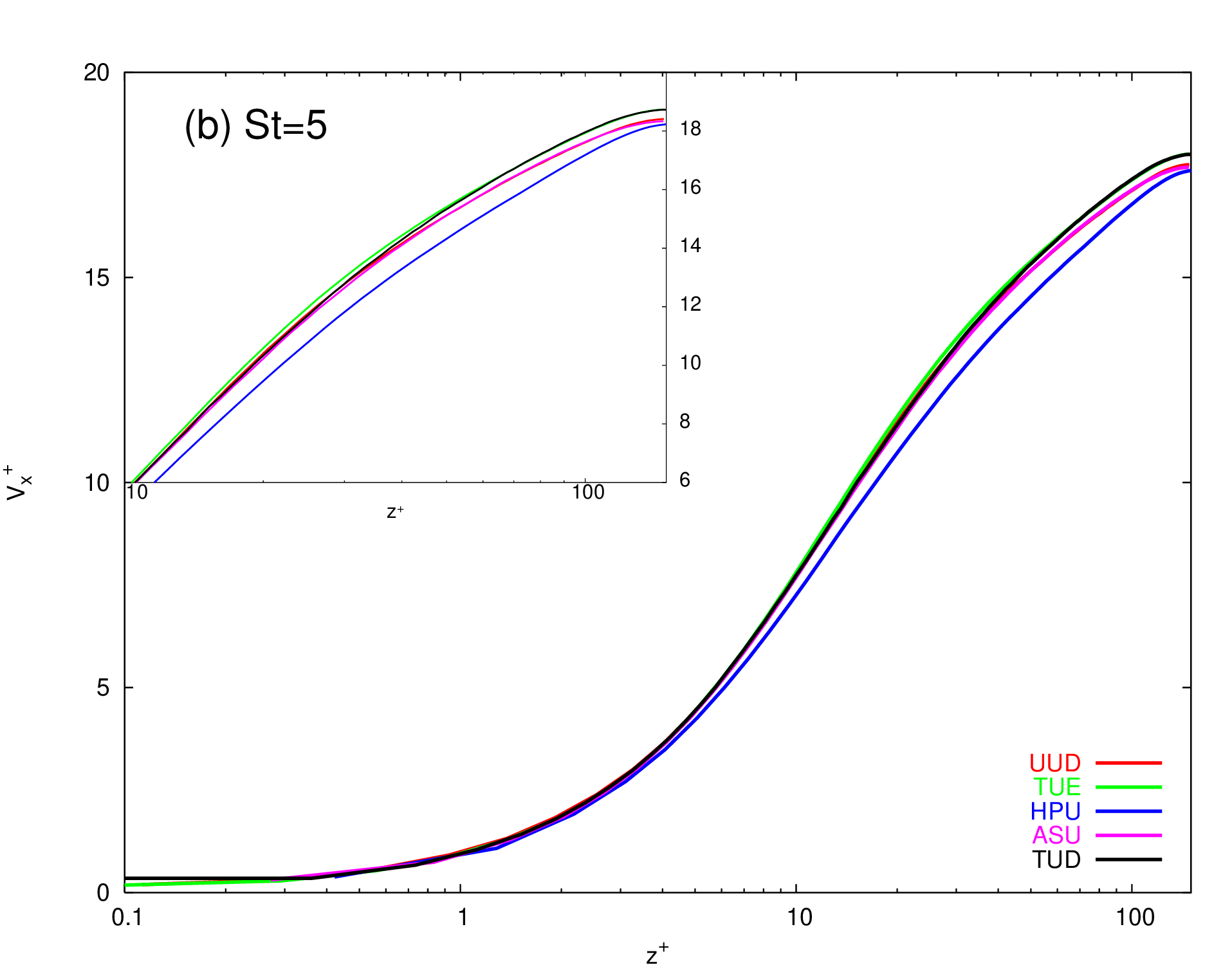}
}
\vspace{-0.5cm}
\centerline{
\includegraphics[width=10.0cm]{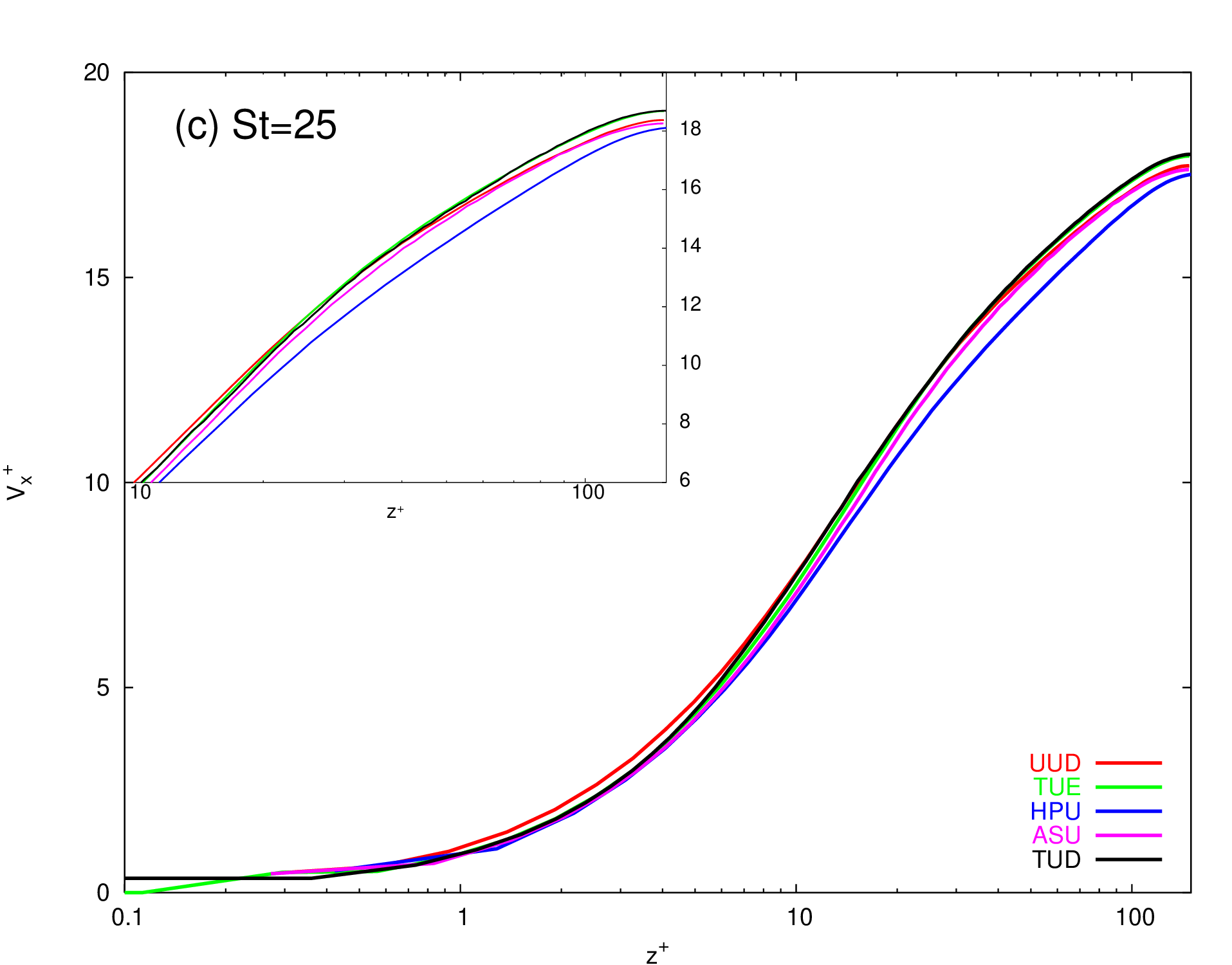}
}
\caption{Mean streamwise particle velocity. (a) $St=1$, (b) $St=5$, (c) $St=25$.}
\label{mean-part-strmws}
\end{figure}

The influence of making different choices in modeling the two
phases is also apparent from particle velocity statistics.
In Fig. \ref{mean-part-strmws} the mean streamwise velocity
profiles, $V_x^+$, for the $St=1$ particles (Fig. \ref{mean-part-strmws}a),
for the $St=5$ particles (Fig. \ref{mean-part-strmws}b) and
for $St=25$ particles (Fig. \ref{mean-part-strmws}c) are
shown.
A close-up view has been included in the diagrams to highlight
the behavior of $V_x^+$ in the outer layer of the channel.
As observed for the fluid velocity, the agreement among
mean quantities is quite satisfactory except for a
velocity deficit in the log-law region.
The profiles of the mean wall-normal velocity are
not shown as they are equal to zero at steady state.
%We remark here that, due to particle drift to the wall,
%particle concentration at the wall will increase continuously
%and the particle wall-normal velocity
%will not be zero until the simulation is fully settled.
%As the stationary state is approached,
%the particle wall-normal velocity will slowly tend to zero
%(as in the present case).
In the database, however, the cross-stream profiles for the mean
relative wall-normal velocity are provided. This velocity, computed
as particle velocity minus fluid velocity {\em seen} by the particles,
can be used to quantify the drift of particles to the wall when the
simulation is not yet fully settled.
%and has the property of not being
%time-dependent (namely, it remains constant during the whole
%deposition process).
%This point should be kept in mind when collecting
%time-averaged particle statistics.

\begin{figure}
\centerline{
\includegraphics[width=10.0cm]{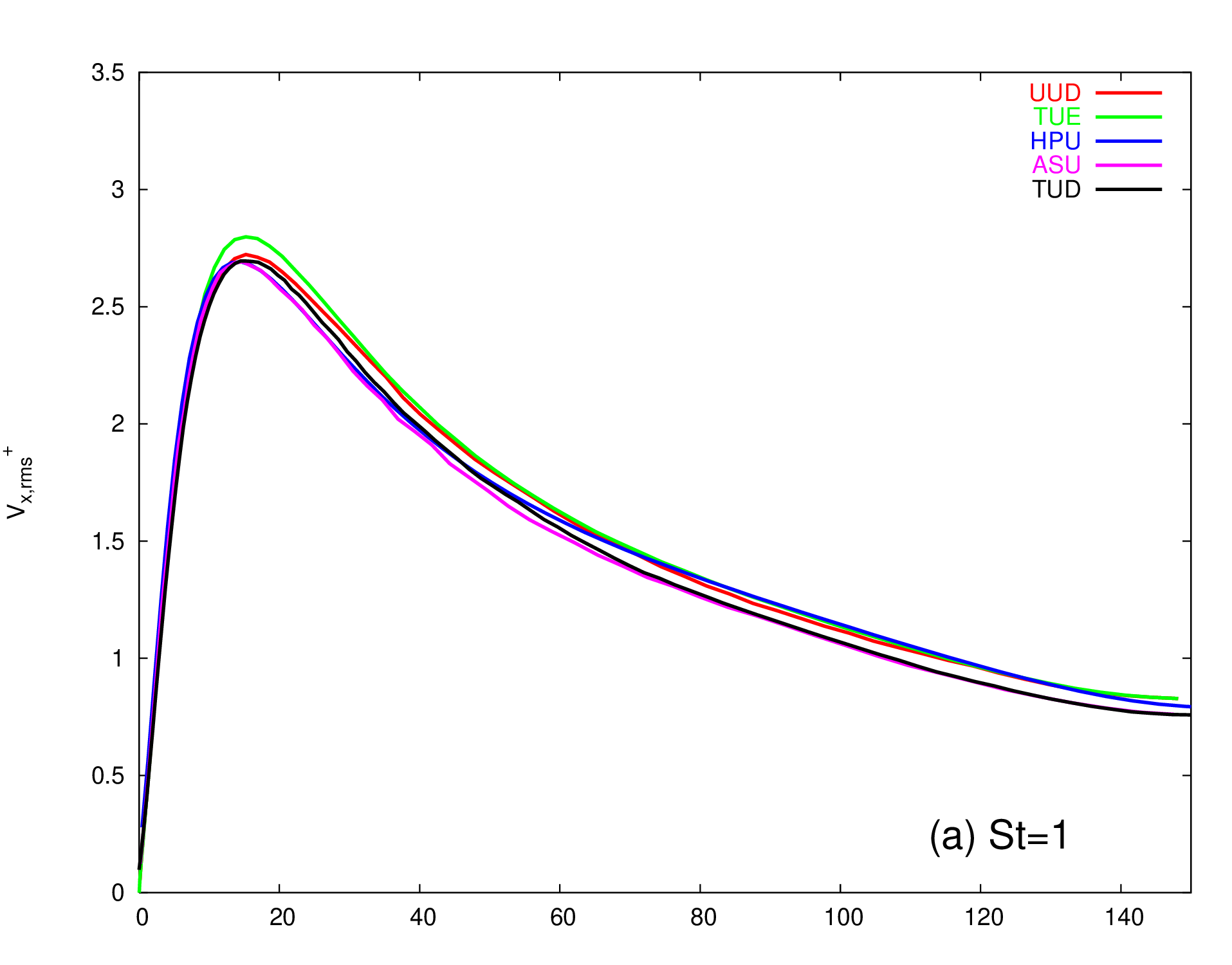}
}
\vspace{-0.8cm}
\centerline{
\includegraphics[width=10.0cm]{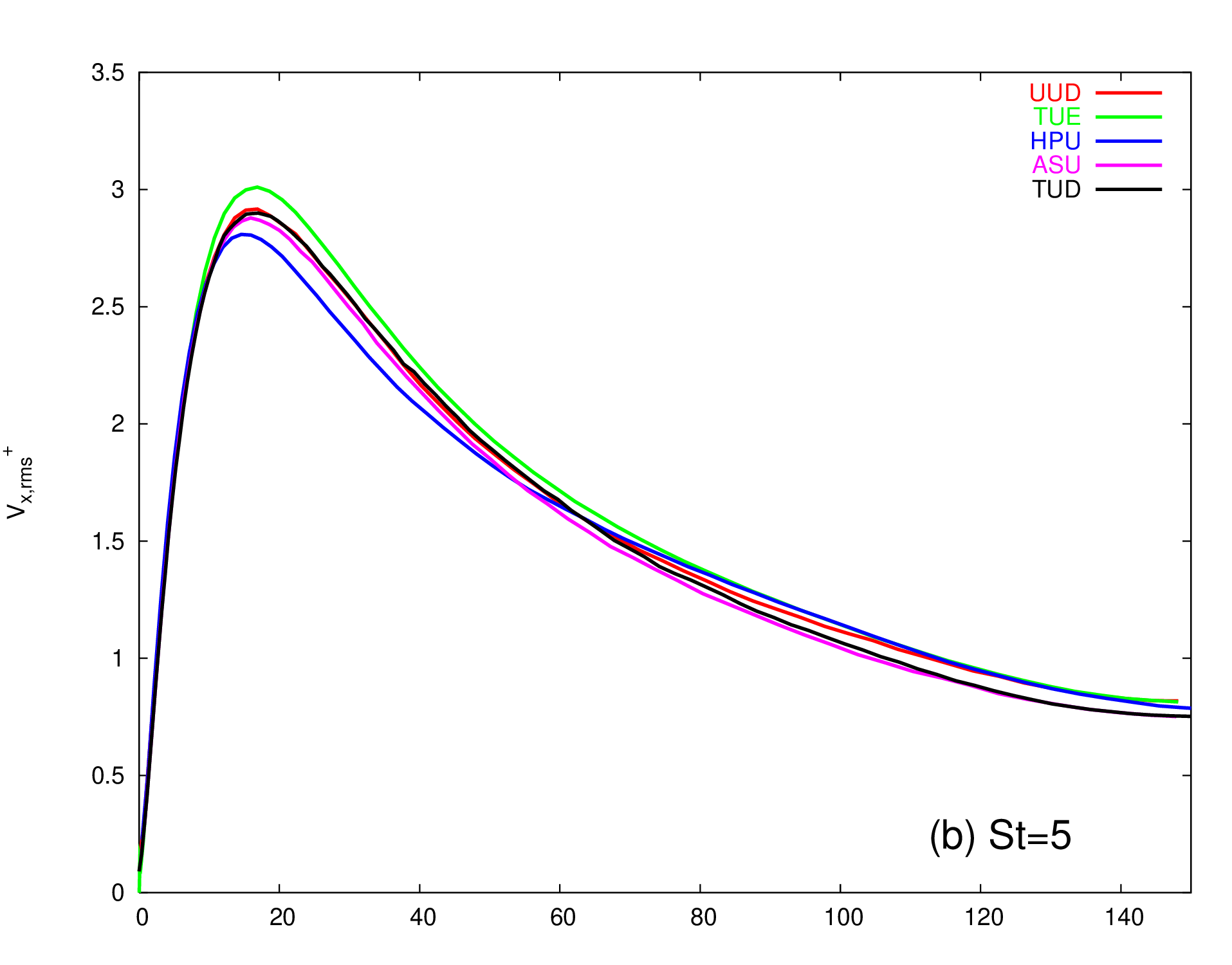}
}
\vspace{-0.8cm}
\centerline{
\includegraphics[width=10.0cm]{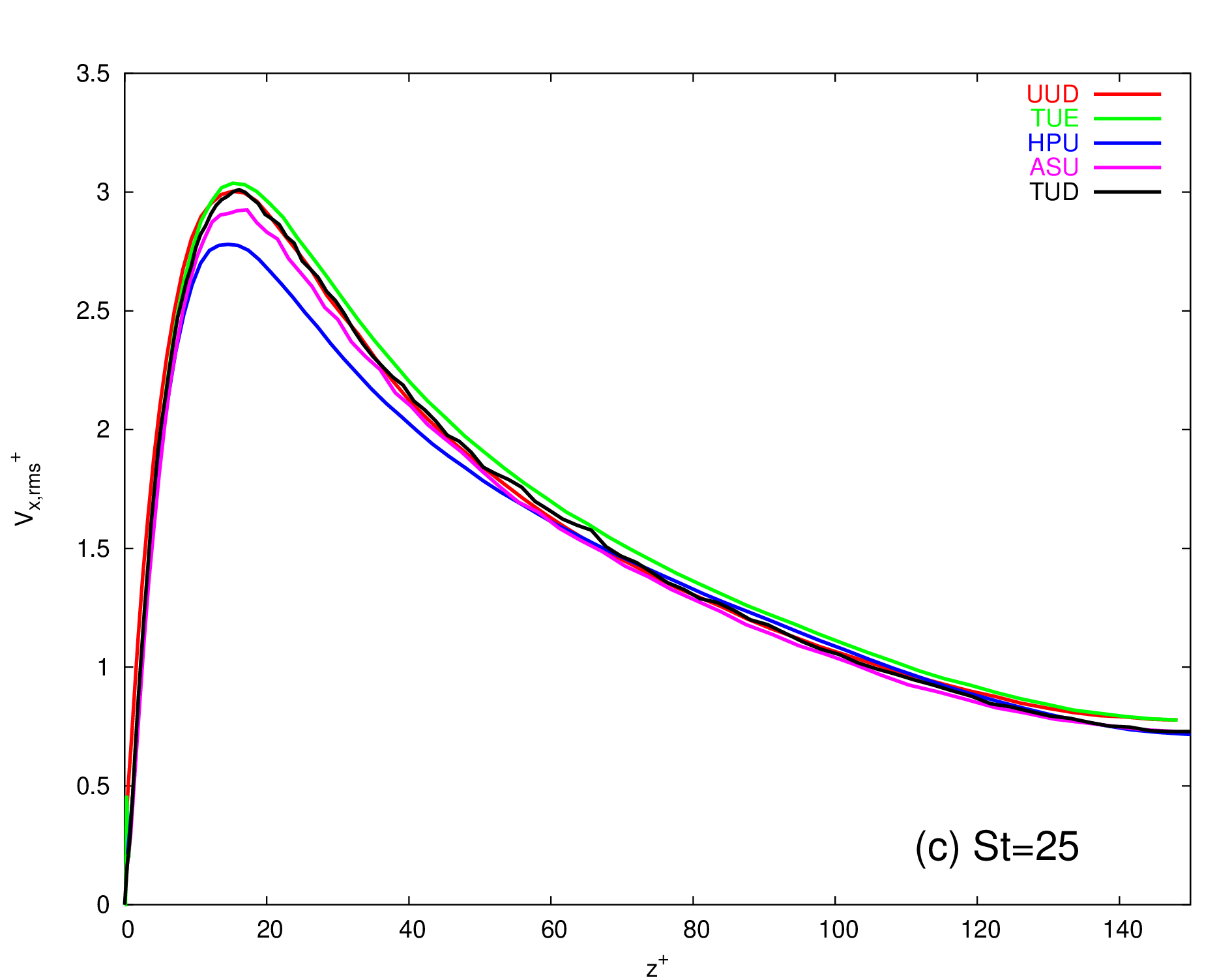}
}
\caption{Root mean square of particle streamwise velocity fluctuations.
(a) $St=1$, (b) $St=5$, (c) $St=25$.}
\label{rms-part-strmws}
\end{figure}

\begin{figure}
\centerline{
\includegraphics[width=10.0cm]{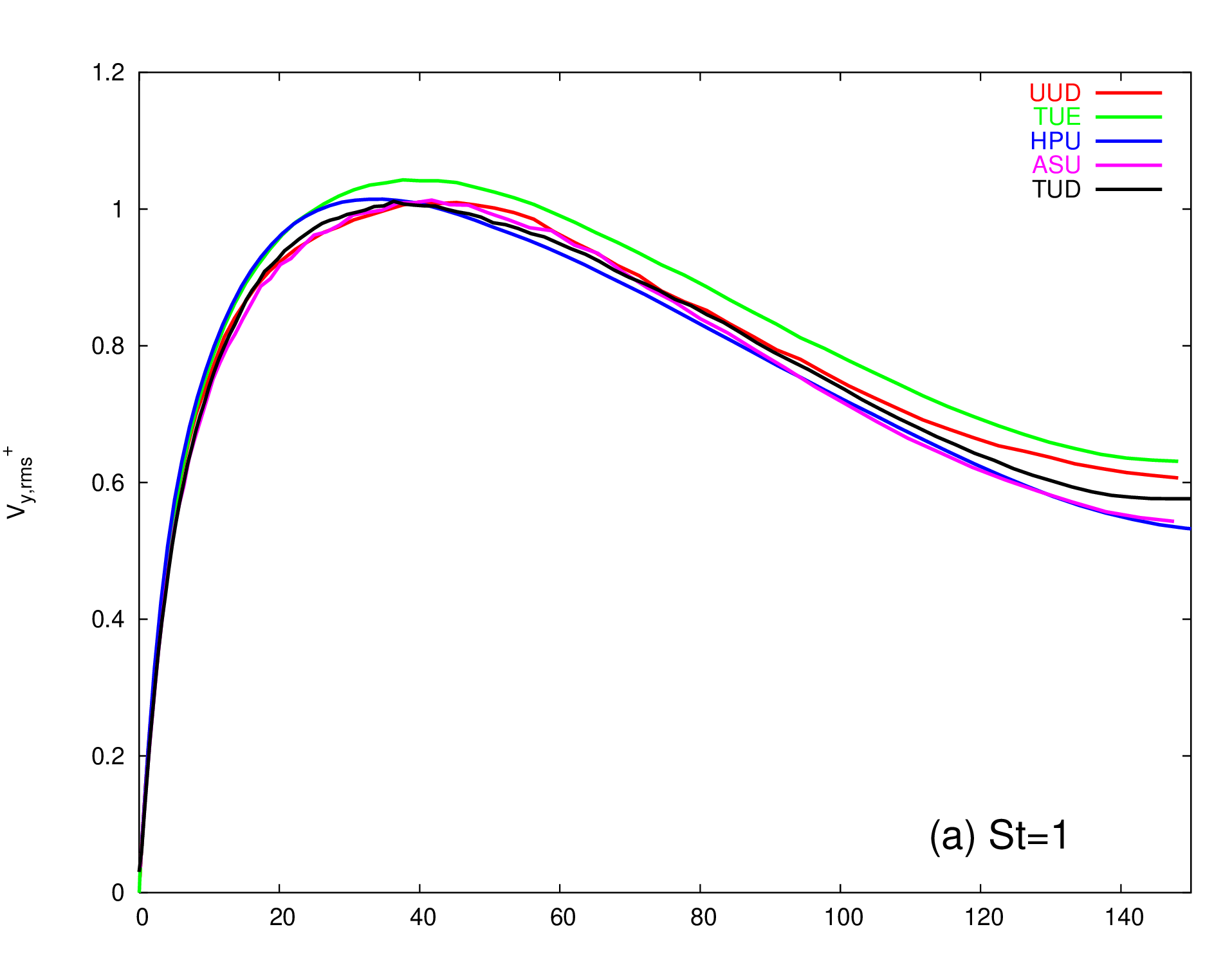}
}
\vspace{-0.8cm}
\centerline{
\includegraphics[width=10.0cm]{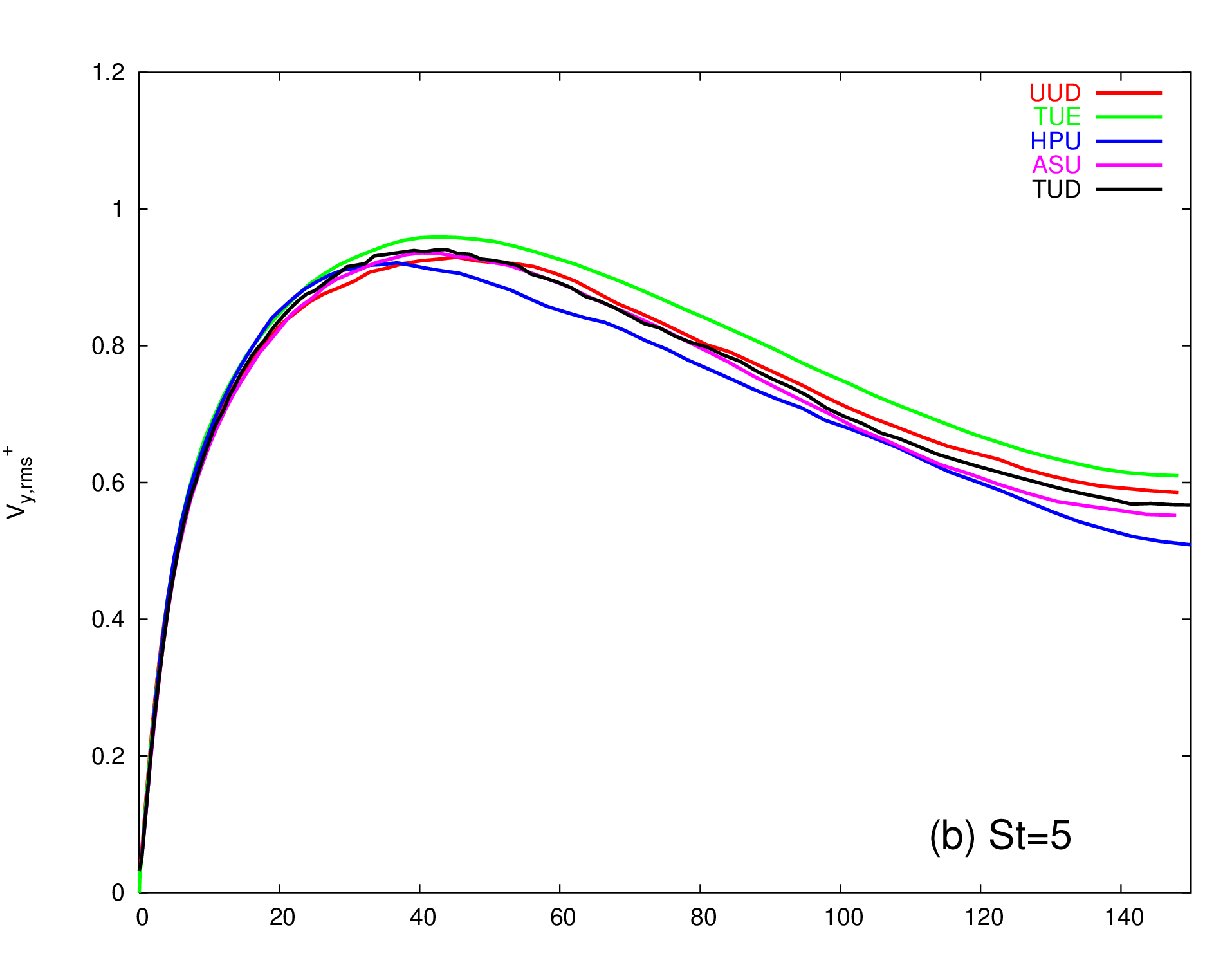}
}
\vspace{-0.8cm}
\centerline{
\includegraphics[width=10.0cm]{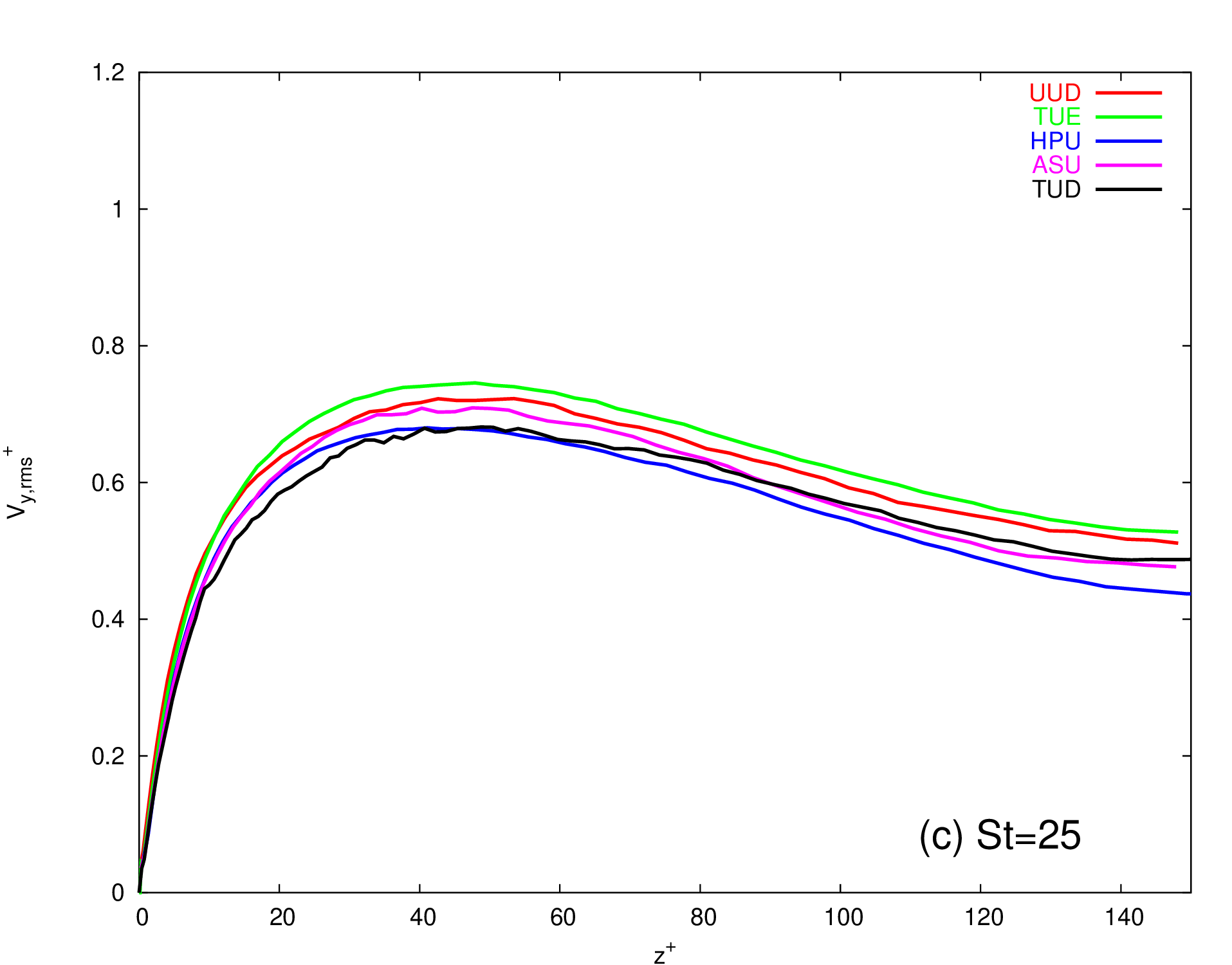}
}
\caption{Root mean square of particle spanwise velocity fluctuations.
(a) $St=1$, (b) $St=5$, (c) $St=25$.}
\label{rms-part-spnws}
\end{figure}

\begin{figure}
\centerline{
\includegraphics[width=10.0cm]{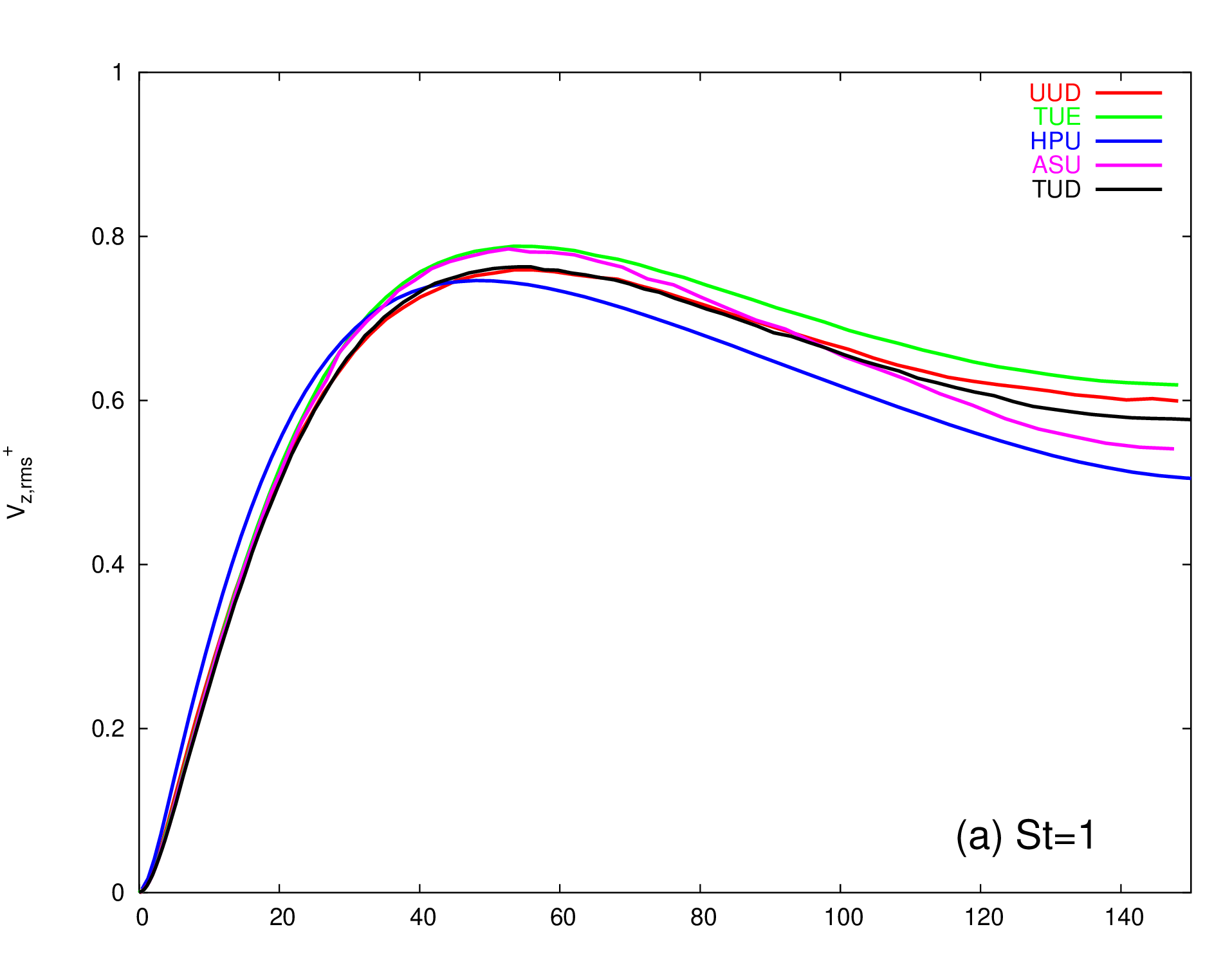}
}
\vspace{-0.8cm}
\centerline{
\includegraphics[width=10.0cm]{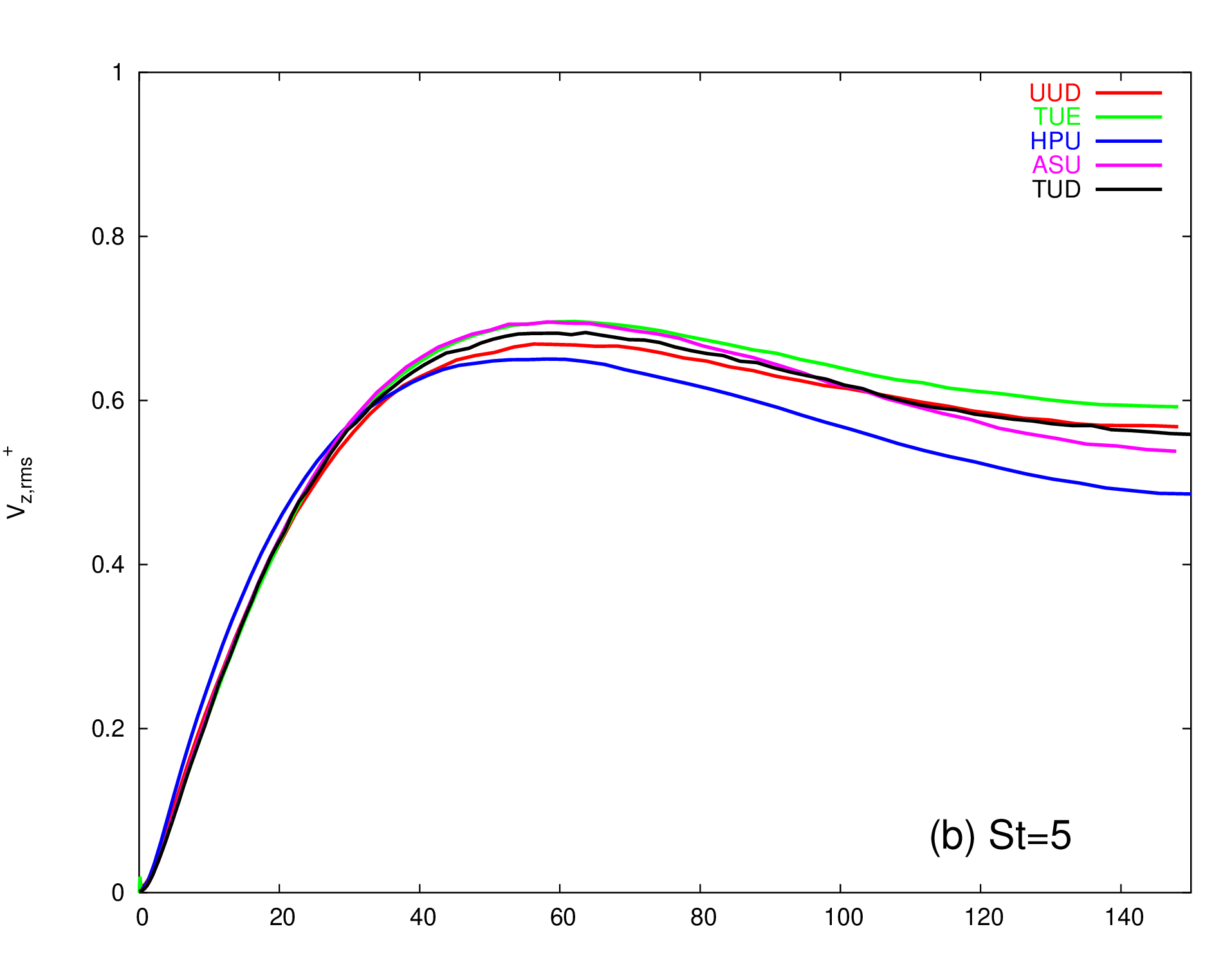}
}
\vspace{-0.8cm}
\centerline{
\includegraphics[width=10.0cm]{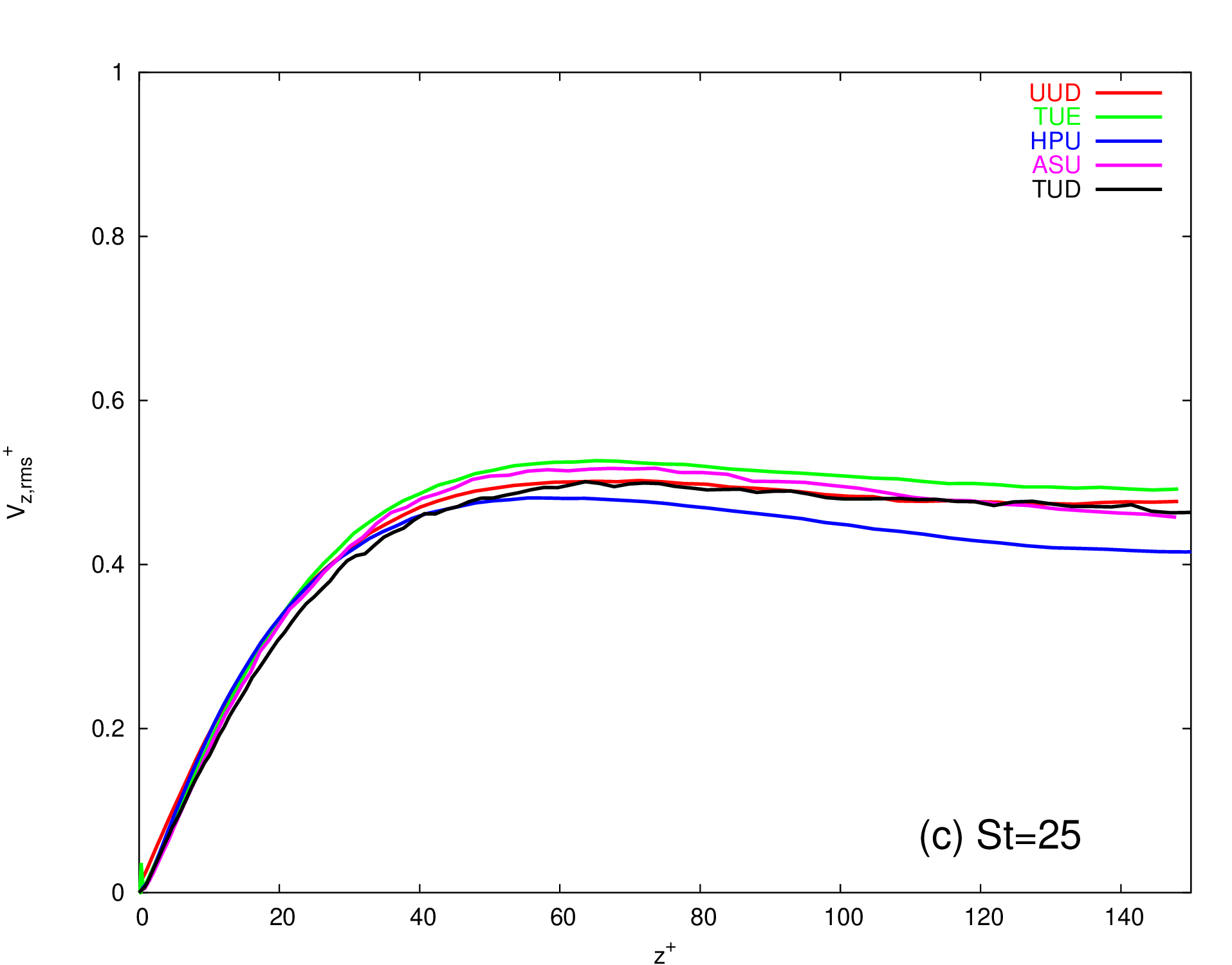}
}
\caption{Root mean square of particle wall-normal velocity fluctuations.
(a) $St=1$, (b) $St=5$, (c) $St=25$.}
\label{rms-part-wall}
\end{figure}

The rms of the particle velocity, $V^+_{i,rms}$,
in the streamwise, spanwise and wall-normal direction
are shown in Fig. \ref{rms-part-strmws}, Fig. \ref{rms-part-spnws}
and in Fig. \ref{rms-part-wall}, respectively.
From Fig. \ref{rms-part-strmws}, it is apparent that the behavior
of $V^+_{x,rms}$ near the centerline is well predicted by all groups,
regardless of the Stokes number.
Near the wall, however, the uncertainty associated with the
calculation of the peak value is higher (even though the peak location
is rather well predicted) and increases with $St$.
In the spanwise (Fig. \ref{rms-part-spnws}) and in the
wall-normal direction (Fig. \ref{rms-part-wall}), the best agreement
between the different groups is found in the near-wall region,
whereas the rms profiles start to deviate from each other as
we move towards the core region of the flow outside
the buffer layer.

\begin{figure}
\centerline{
\includegraphics[width=10.0cm]{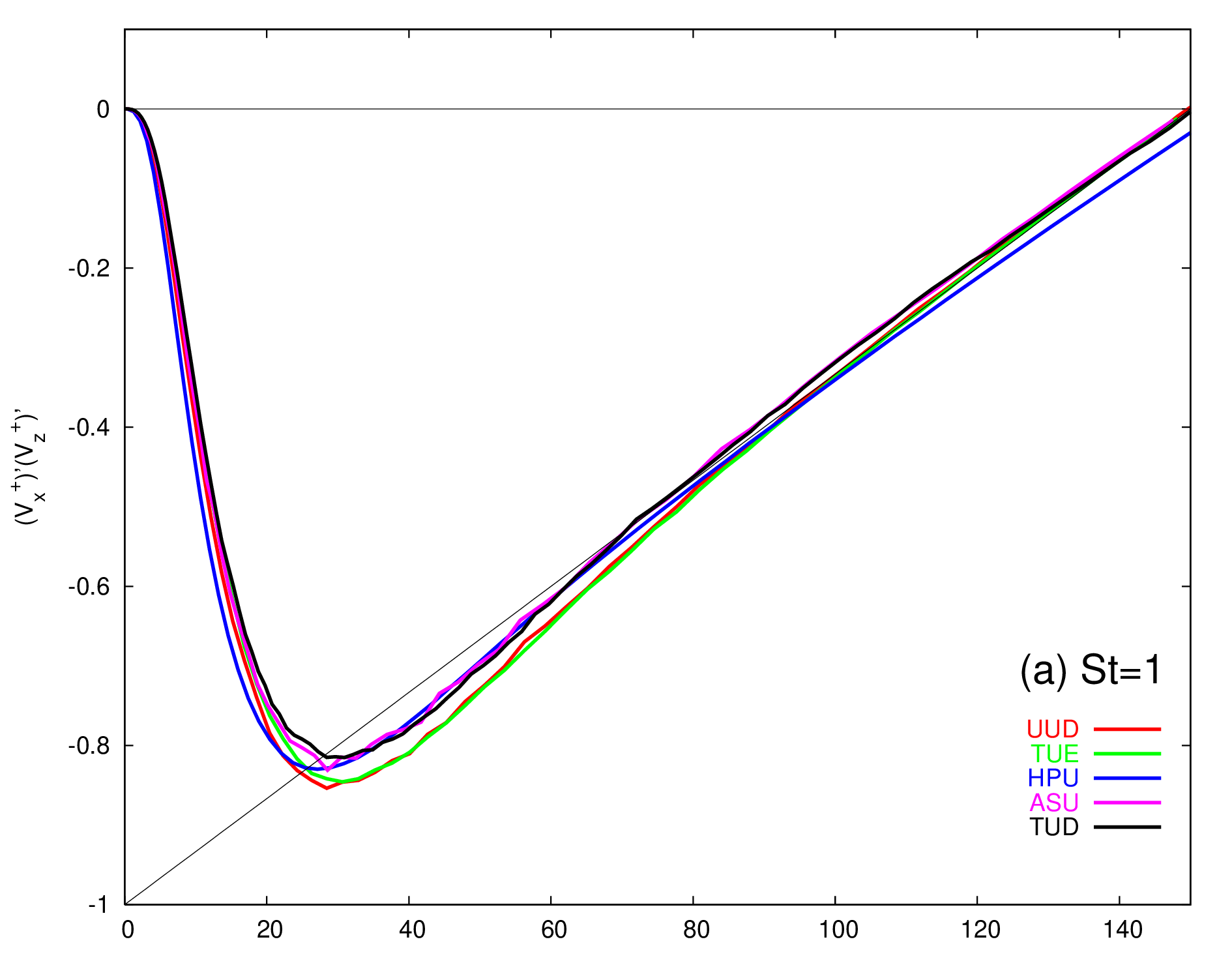}
}
\vspace{-0.5cm}
\centerline{
\includegraphics[width=10.0cm]{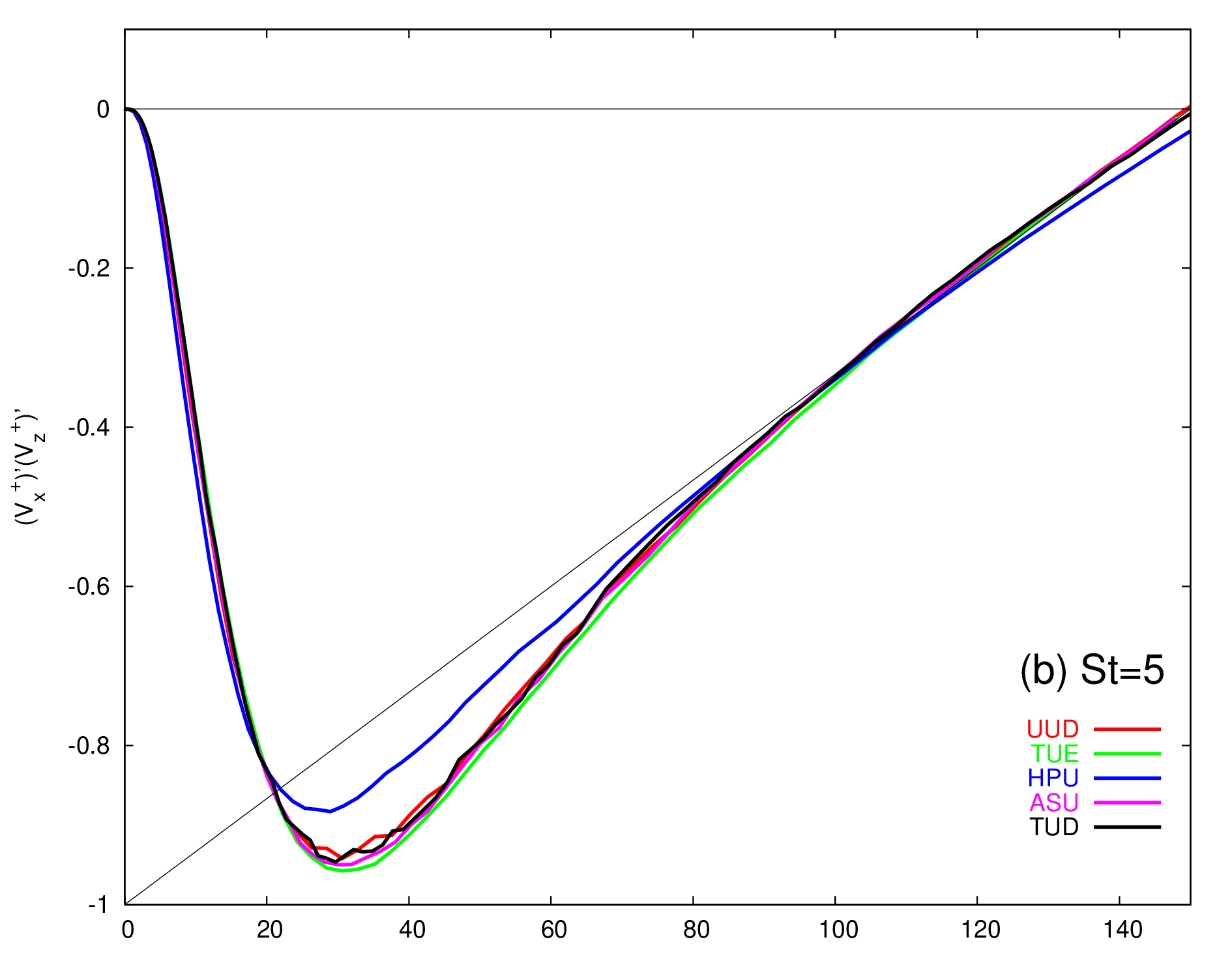}
}
\vspace{-0.5cm}
\centerline{
\includegraphics[width=10.0cm]{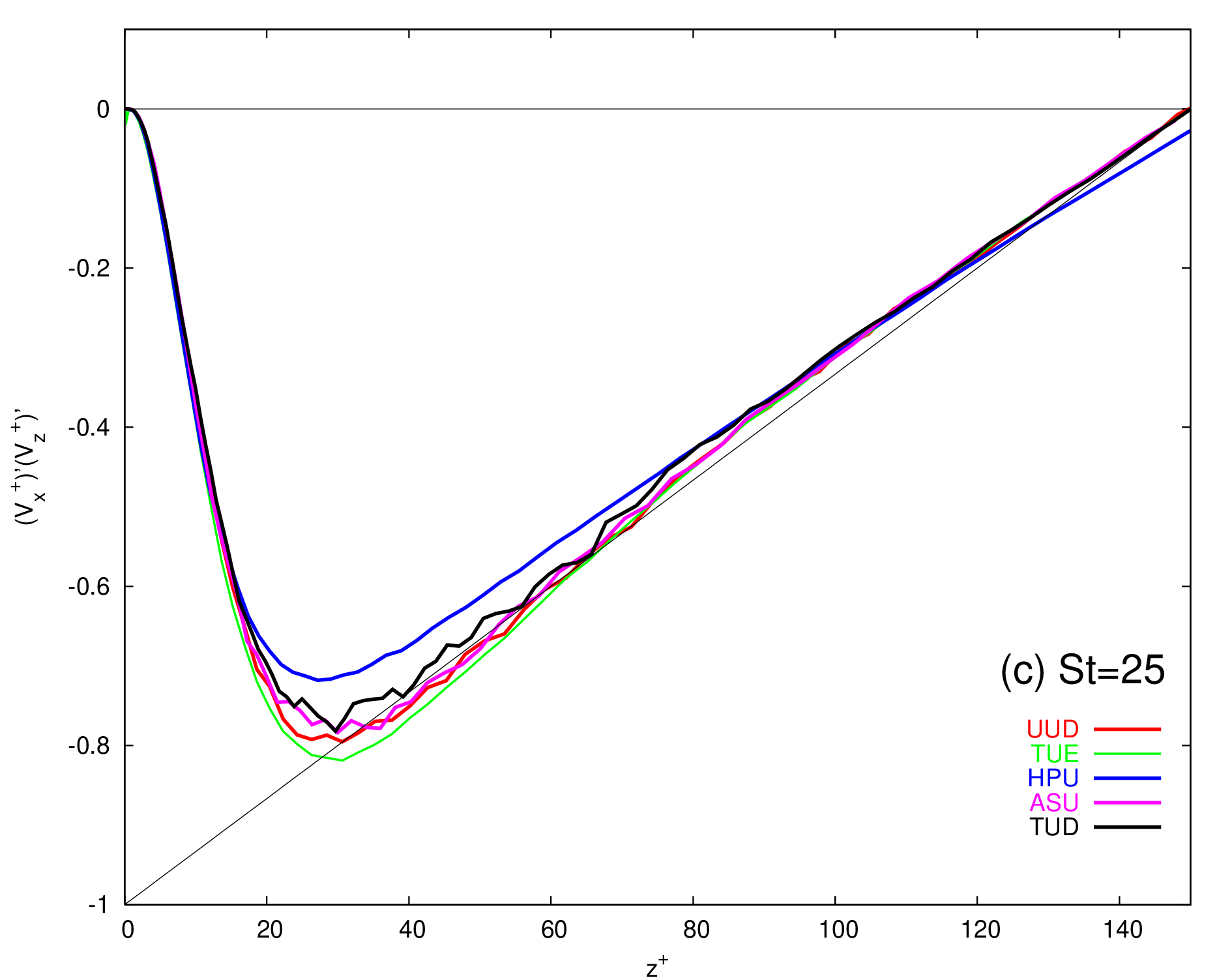}
}
\caption{Particle Reynolds stresses: $(V_x^+)'(V_z^+)'$ component.
(a) $St=1$, (b) $St=5$, (c) $St=25$.}
\label{stresses}
\end{figure}

Analysis of the mean and rms values of particle velocity seems to indicate
that single-point particle velocity statistics are not much affected by
the different predictions of particle concentration. To corroborate this
conclusion, in Fig.  \ref{stresses} we show the $(V_x^+)'(V_z^+)'$ component
of the Reynolds stress tensor for the particles. Detailed
knowledge of the elements of this tensor is crucial to validate
theoretical models of particle deposition in wall-bounded turbulent
flows that try to reproduce the convective wallward drift of particles by
assuming local equilibrium between the particles and the fluid
turbulence (see the deposition model by Young and Leeming (1997),
for instance).
The trend we can observe from Fig. \ref{stresses} is similar to
that observed for the rms of the particle velocities: there is a
good agreement for the smaller particles and an increasing uncertainty
associated with the calculation of the peak value for the larger particles.
In particular, the profile of Group HPU is characterized by a
smaller absolute value of the peak for both the $St=5$ and the
$St=25$ particles
which is likely related to an underprediction of the particle
velocity fluctuations in the near-wall region as compared with
the other groups.
In Fig. \ref{stresses} (and in Figs. \ref{rms-part-strmws}
to \ref{rms-part-wall} as well)
some profiles appear to be a little bit ragged
due to smaller intervals chosen for time averaging.
The length of the averaging interval was one of the parameters
that could be chosen by the participants.
%in this way, it can be seen how the accuracy of the results depends on the choice made.
Here, a longer time span would have certainly smoothed out the
profiles but it would have not changed their relative position within
the chart.

One interesting aspect of the test case is that it gives
the chance to single out the effect of each possible
source of error on the observed results through a well-aimed
parametric study. This study is not currently
available (as it is beyond the scope of the benchmark
calculation) but can be definitely regarded as a possible
extension to the base simulation.

%%%%%%%%%%%%%%%%%%%%%
\section{Conclusions}
%%%%%%%%%%%%%%%%%%%%%

The dispersion of particles with finite inertia
in wall-bounded turbulent flows is
of fundamental importance for numerous applications
in industry and environment. The dispersion process, however,
is characterized by
complex phenomena such as non-homogeneous distribution, large-scale
clustering and preferential concentration
in the near-wall region due to the inertial bias between the denser
particles and the lighter surrounding fluid (Marchioli and Soldati, 2002;
Eaton and Fessler, 1994).
Direct Numerical Simulation, even at moderate Reynolds number,
coupled with Lagrangian Particle Tracking has been widely used to study
these macroscopic phenomena, for instance in vertical turbulent pipe
(Vreman, 2007; Uijttewaal and Oliemans, 1996)
and channel flows (Li et al., 2001; McLaughlin, 1989),
%\cite{Marchioli2002, Marchioli2004}.
and represents a useful tool to provide physical
insights, new modeling ideas and benchmark cases (Moin \& Mahesh,
1998; Yeung, 2002).

In this paper, we have presented the main results produced by an
international collaborative test case in which direct comparison
is made among the numerical predictions obtained by different computational
codes for the common problem of turbulent particle dispersion in channel
flow.
%This flow configuration has been chosen since it represents a well-known
%instance of wall-bounded flow.
%This benchmark test has produced a DNS statistical database
%for the specific benchmark case of dilute particle-laden
%turbulent channel flow at $Re_{\tau}=150$.
A comprehensive database of statistics for the fluid and for the
particles has been gathered and made available in the form of
post-processed ASCII files at {\sc{http://cfd.cineca.it/cfd/repository/}}.
In addition to repository files containing the statistical datasets,
the database includes raw data for the instantaneous fully-developed flow field
and for the particle position/velocity:
these data are made available in the
form of formatted ASCII files to users who need to compute
specific statistics not yet included in the database.
Another original (and very important) feature of the database
is that it was collected under statistically-steady condition
for the particle distribution:
the database may thus bring significant advantages both
from the computational viewpoint (starting a new simulation with
steady-state initial condition for the dispersed phase may allow
large savings in terms of CPU time) and from the modeling viewpoint
(the datasets may be used to validate closure approximations
for models based on the assumption that the flow of the particle phase
is fully developed: see Sergeev et al. (2002), for instance).

On the basis of the results discussed in this paper, the
conclusions listed below can be drawn.
\begin{itemize}
\item The database represents a homogeneous source of data
on DNS and LPT not previously available that can be used as
a benchmark to test the performance of new numerical methods
or as a tool to validate theoretical models for
the gas-solid interactions in channel flow (for instance,
models including a-posteriori Large-Eddy Simulations).
\item Direct comparison of the statistics allows clearcut
observation of {\em (i)} how different codes perform when applied to
the same problem with a well-defined simulation setting
and of {\em (ii)} how the accuracy of the results depends
on the choices made in terms of simulation parameter values.
As demonstrated by the several previous papers published
independently by each of the participating groups (see Marchioli and Soldati, 2002;
Portela and Oliemans, 2003; Kuerten, 2006; Arcen et al., 2006 for instance)
all methods have been used to produce
DNS-quality data to investigate on the physics of turbulent particle dispersion
in wall-bounded flows and/or to benchmark simpler models.
However, direct comparison of the results brings to the following
{\em caveat}: even when
the most accurate numerical tools are used and all the
simulation requirements are fulfilled, one may find non-negligible
quantitative differences in the statistics.
Of course, there will be a {\em best} prediction for
a given statistical quantity, yet such best prediction is not known a priori.
For this reason, it appears very difficult to
provide clearcut indications on the reliability of each
dataset: we can just observe that the range
of wall concentration predictions can be accepted as a good measure
of particle wall concentration under the modelling assumptions used
in this work.
\item parametric studies performed apart from the base simulations
are required to single out the effect of changing one simulation
parameter (or more) from the macroscopic particle behavior.
\end{itemize}

The test case calculations can be regarded as a challenge
to approach more complex problems in two-phase flow predictions
and will hopefully stimulate further improvements and developments
of numerical methods and models.
To this aim, test case calculations will be continued by extending
the base simulation presented here. Specifically, all participant
groups will include one or more additional simulation parameters.
To compare results more easily, the choice will be restricted to
parameters dealing with the physical modeling
of the flow, such as fluid-particle two-way coupling, inter-particle
collisions, lift force models and sub-grid scale effects on
particle motion in Large-Eddy Simulation fields.
Further parameter analysis will be planned at a later stage and
other statistical quantities will be made available as they
are extracted from the simulations.

%%%%%%%%%%%%%%%%%%%%%%%%%%
\section*{Acknowledgments}
%%%%%%%%%%%%%%%%%%%%%%%%%%

We wish to thank the Cineca supercomputing center (Bologna, Italy)
for the hosting of the DNS database.

\vspace*{0.8\baselineskip}
\noindent{REFERENCES}

%\include{captions}

%
% TABLE 1
%
\begin{table}[hb]
\begin{small}
\begin{center}
\begin{tabular}{c c c c c c}
\hline
$St = \tau_{p}^{+}$ & $\tau_{p}~(s)$ & $d_{p}^{+}$ & $d_{p}$ (${\mu}m$) & $V_{s}^{+}=g^+ St$  & $Re_{p}^{+}=\frac{V_s^+ d_{p}^{+}}{\nu^+}$\\ \hline \hline
%0.2 & $2.265 \cdot 10^{-4}$ & $0.068$ & $9.12$ & $0.019$ & $1.39 \cdot 10^{-3}$\\
1   & $1.133 \cdot 10^{-3}$ & $0.153$ & $20.4$ & $0.094$ & $0.01444$\\
5   & $5.660 \cdot 10^{-3}$ & $0.342$ & $45.6$ & $0.472$ & $0.16127$\\
25  & $2.832 \cdot 10^{-2}$ & $0.765$ & $102$ & $2.360$ & $1.80505$\\
%125 & $1.415 \cdot 10^{-1}$ & $1.71$  & $228$ & $11.79$ & $20.1585$\\
\hline
\end{tabular}
\vspace{0.3cm}
\caption{\label{part} Particle parameters.}
\end{center}
\end{small}
\end{table}

%
% TABLE 2
%
\begin{table}[hb]
\begin{small}
\begin{center}
\begin{tabular}{c c c c c c c}
\hline
Group & Flow   & Time integration     & Time         & Fluid         & Grid       & Wall-normal \\
      & solver & of fluid (non-linear & integration  & velocity      & resolution & distribution of \\
      &        & + viscous terms)     & of particles & interpolation &            & collocation points\\ \hline \hline
UUD & PS  & AB2 + CN & RK4 ($\Delta t^+ = 0.45$) & L6  & $128 \times 128 \times 129$ & Chebyschev \\
TUE & PS  & RK3 + CN & H2  ($\Delta t^+ = 0.032$) & LH4 & $128 \times 128 \times 129$ & Chebyschev \\
HPU & FD2 & RK3 + CN & RK3 ($\Delta t^+ = 0.05$) & H3  & $192 \times 160 \times 128$ & HT (SF=1.7)\\
ASU & FD2 & AB2 + CN & AB2 ($\Delta t^+ = 0.05$) & H3  & $128 \times 128 \times 129$ & HT (SF=1.7)\\
TUD & FV2 & AB2      & RK2 ($\Delta t^+ = 0.026$) & TL  & $192 \times 192 \times 192$ & HT (SF=1.7) \\
\hline
\end{tabular}
\vspace{0.3cm}
\caption{
Summary of numerical methodologies. Nomenclature used
in this table is as follows:
{\em (i) Flow Solver} -
PS: Pseudo-Spectral,
FD2: $2^{nd}$ order Finite Differences,
%CD2: $2^{nd}$ order Centered Differences,
FV2: $2^{nd}$ order Finite Volumes;
{\em (ii) Time Integration} - 
AB2: $2^{nd}$ order Adams-Bashforth,
CN: implicit Crank-Nicolson,
RK2: $2^{nd}$ order Runge-Kutta,
RK3: $3^{rd}$ order Runge-Kutta,
RK4: $4^{th}$ order Runge-Kutta,
H2: $2^{nd}$ order Heun method,
$\Delta t^+$: non-dimensional time step size;
{\em (iii) Fluid Velocity Interpolation} -
L6: $6^{th}$ order Lagrange polynomials,
LH4: $4^{th}$ order Lagrange-Hermite polynomials,
H3: $3^{rd}$ order Hermite polynomials.
TL: Tri-Linear;
{\em (iv) Wall-Normal Collocation Points} -
HT: Hyperbolic Tangent
(SF: Stretching Factor).
}
\label{dns}
\end{center}
\end{small}
\end{table}

%
% Fig. 1
%

\clearpage
\newpage

\clearpage
\newpage

%
% Fig. 2
%

\clearpage
\newpage

%
% Fig. 3
%

%
% Fig. 4
%

\clearpage
\newpage

%
% Fig. 4
%

\clearpage
\newpage

\clearpage
\newpage

%
% Fig. 5
%

%
% Fig. 6
%

\clearpage
\newpage

%
% Fig. 7
%

\clearpage
\newpage

%
% Fig. 8
%

\clearpage
\newpage

%
% Fig. 10
%

\clearpage
\newpage

%
% Fig. 9
%

\clearpage
\newpage

\end{document}